\documentclass[reprint,amsmath,amssymb,longbibliography,superscriptaddress,
 aps,prl]{revtex4-2} 
\usepackage{hyperref}
\usepackage{graphicx}
\usepackage{amsmath}
\usepackage{color}
\usepackage{physics}
\usepackage[T1]{fontenc}
\usepackage{pdfpages} 
\usepackage{pgffor} 
\usepackage{multirow}
\usepackage{lineno}
\usepackage[normalem]{ulem}

\makeatletter
\AtBeginDocument{\let\LS@rot\@undefined}
\makeatother

\def\supplementfilename{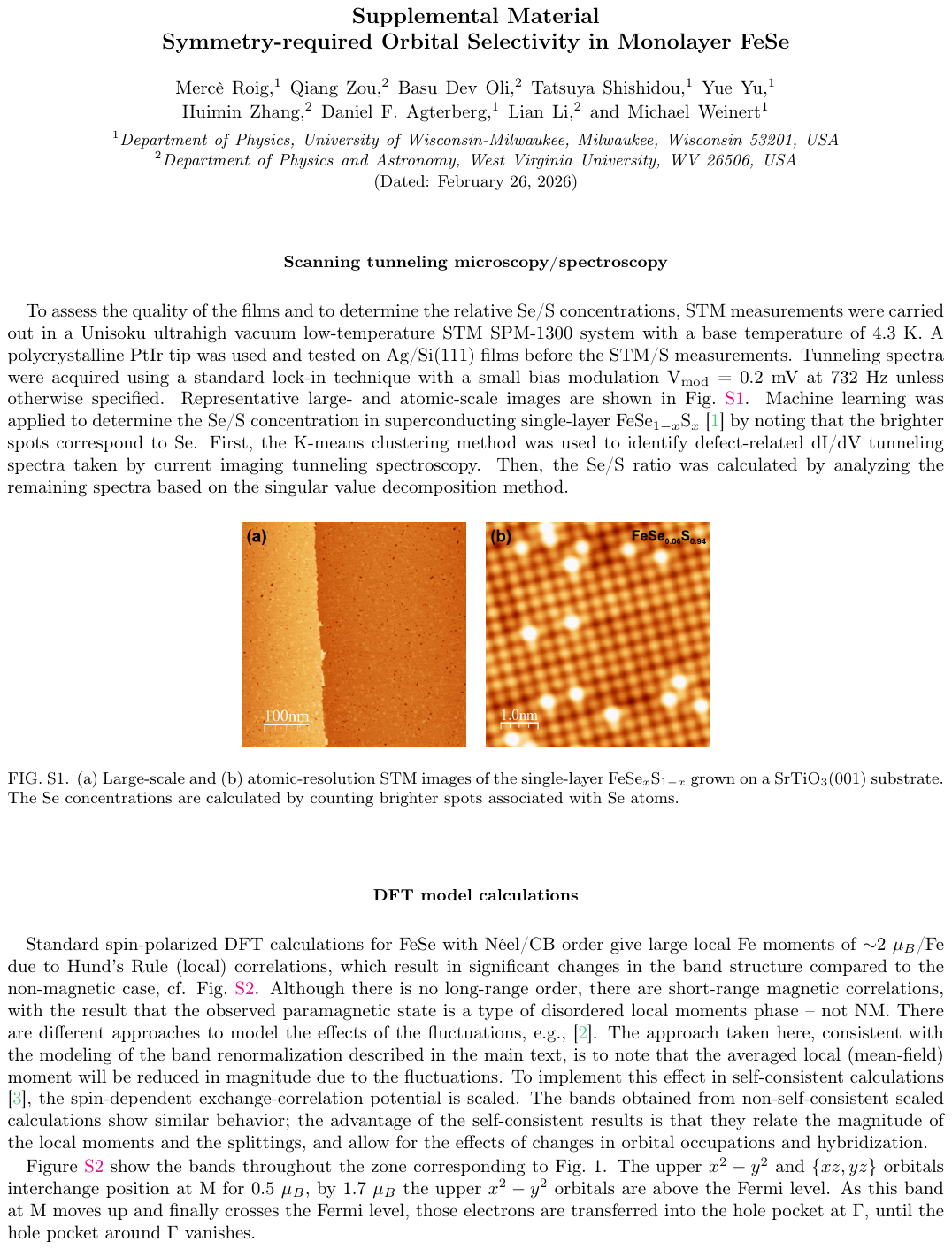}
\def\numbersupplementpages{\the\pdflastximagepages}

\newif\ifarXiv
\arXivtrue 

\definecolor{limegreen}{rgb}{0.2, 0.8, 0.2}
\definecolor{orange}{rgb}{1.0, 0.5, 0.0}
\definecolor{blue(ncs)}{rgb}{0.0, 0.53, 0.74}

\definecolor{emerald}{rgb}{0.31, 0.78, 0.47}

\hypersetup{%
     colorlinks=true,
     linkcolor=magenta,
     filecolor=blue,
     citecolor=emerald,      
     urlcolor =blue(ncs),
}

\newcommand{\kv}{{\bf k}}
\newcommand{\qv}{{\bf q}}

\begin{document}
\title{Symmetry-required Orbital Selectivity in Monolayer FeSe}

\author{Mercè Roig}
\altaffiliation{These authors contributed equally to this work.}
\affiliation{Department of Physics, University of Wisconsin-Milwaukee, Milwaukee, Wisconsin 53201, USA}

\author{Qiang Zou}
\altaffiliation{These authors contributed equally to this work.}
\affiliation{Department of Physics and Astronomy, West Virginia University, WV 26506, USA}

\author{Basu Dev Oli}
\affiliation{Department of Physics and Astronomy, West Virginia University, WV 26506, USA}

\author{Tatsuya Shishidou}
\affiliation{Department of Physics, University of Wisconsin-Milwaukee, Milwaukee, Wisconsin 53201, USA}

\author{Yue Yu}
\affiliation{Department of Physics, University of Wisconsin-Milwaukee, Milwaukee, Wisconsin 53201, USA}

\author{Huimin Zhang}
\affiliation{Department of Physics and Astronomy, West Virginia University, WV 26506, USA}

\author{Daniel F. Agterberg}
\affiliation{Department of Physics, University of Wisconsin-Milwaukee, Milwaukee, Wisconsin 53201, USA}

\author{Lian Li}
\affiliation{Department of Physics and Astronomy, West Virginia University, WV 26506, USA}

\author{Michael Weinert}
\affiliation{Department of Physics, University of Wisconsin-Milwaukee, Milwaukee, Wisconsin 53201, USA}

\begin{abstract}
Orbital-selective correlations have been observed to play an important role in Fe-based superconductors.  Here, in contrast to previous site-local Mott transition-based origins, we present a band-theory-based mechanism for orbital-selective physics in monolayer FeSe, for which only electron pockets appear. Underlying our mechanism is the observation in density functional theory (DFT) calculations that around the M point in the Brillouin zone, antiferromagnetic fluctuations are strongly coupled to electrons in $x^2-y^2$ orbitals but weakly coupled to those in $\{xz,yz\}$ orbitals. 
Symmetry-arguments reveal that this orbital selective coupling originates from the different intertwined orbital and Fe-site sublattice Bloch wavefunctions for these two sets of orbitals at the M point, specifically, the $x^2-y^2$ orbitals can be Fe-site localized. The strong coupling of electrons in $x^2-y^2$ orbitals to the magnetic fluctuations enables orbital-selective electronic renormalizations that can account for important features of our angle-resolved photoemission spectroscopy (ARPES) measurements. Our symmetry-required mechanism for orbital selective physics can be generalized to a range of crystal space groups with four-fold and six-fold screw axes. 
\end{abstract}

\date{\today}%

\maketitle

\section{Introduction}
Orbital selective Mott physics, in which electrons in different atomic orbitals experience different Mott localization transitions, often forms the basis for strong-coupling interpretations of single-particle electronic orbital-dependent renormalizations. Indeed, this point of view has been widely adopted in Fe-based superconductors, where an orbital-dependent Mott transition has been proposed to explain the enhanced correlation effects observed for electrons in $x^2-y^2$ orbitals compared to $\{xz,yz\}$
orbitals~\cite{Sprau2017Jul,Kreisel2017May,Liu2022Apr,Kostin2018Oct,Yi2015Jul,De'Medici2014Apr,Huang2022Jan}. Here, electrons in $x^2-y^2$ orbitals are believed to be closer to a Mott transition than those in $\{xz,yz\}$ orbitals, in part because $\{xz,yz\}$ orbitals have higher degeneracy in tetragonal materials~\cite{De'Medici2014Apr}. Fe-based superconductors are typically characterized as having intermediate electronic correlations, with similar magnitudes for the bandwidth and Coulomb repulsion~\cite{fphys21}. It is natural to ask whether orbital selective physics can also arise from a mechanism originating from the electronic band structure. 
Indeed, in Kagome materials, band structure-driven sublattice interference drives a correlated broken time-reversal charge density state~\cite{Kiesel:2012,Kiesel:2013}, which is consistent with experimental observations~\cite{Mielke:2022}. This sublattice interference is a consequence of Bloch states near Van-Hove singularities composed of intertwined $d$-orbitals and Kagome sublattice sites. 

\begin{figure}[t]
\begin{center}
\includegraphics[angle=0,width=1.0\columnwidth]{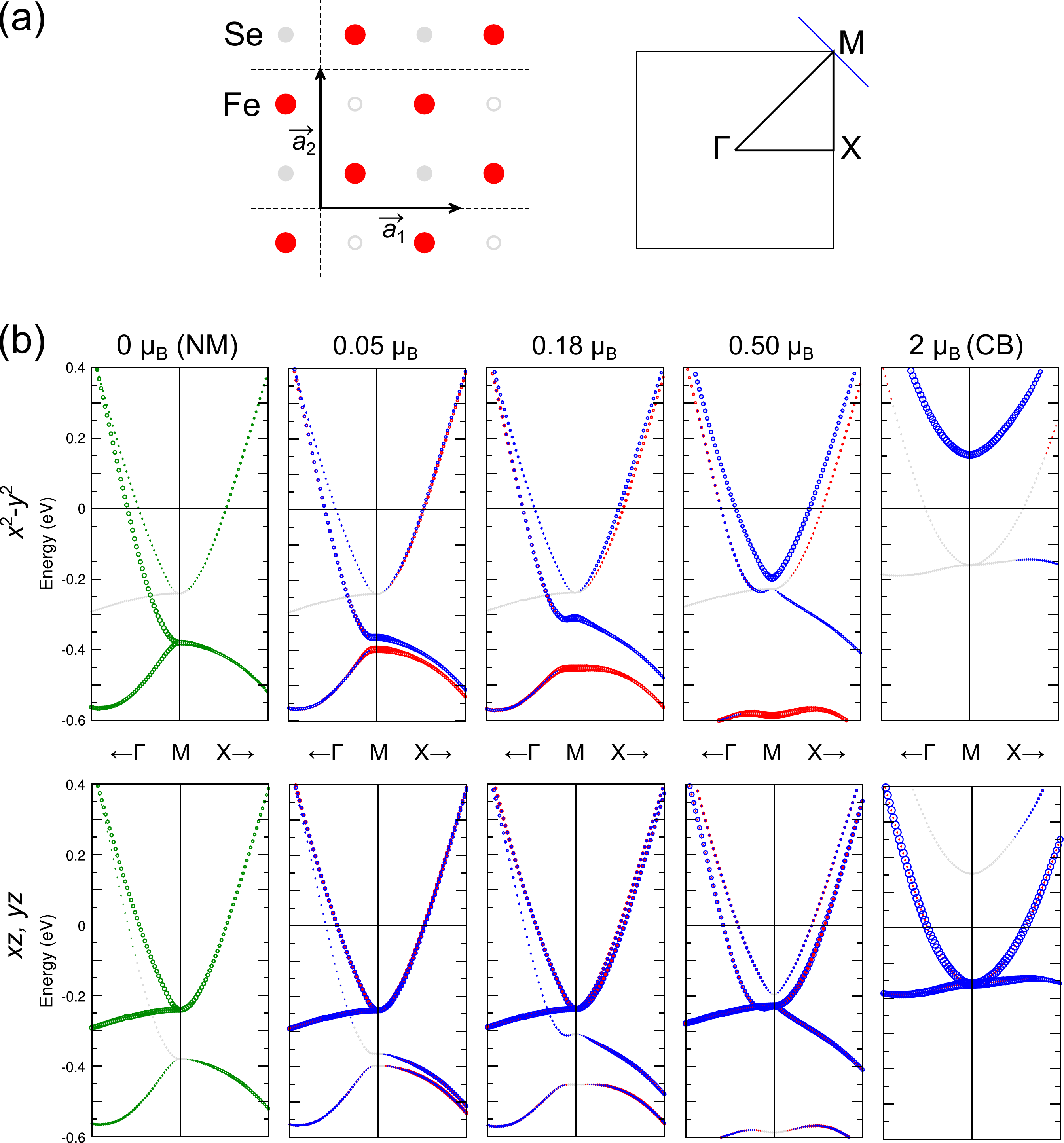}
\caption{(a) Schematics of the (2 Fe unit cell) crystal structure and Brillouin zone.  The Se atoms above (below) the Fe plane are denoted by filled (open) gray circles. The blue line indicates the cut for the ARPES data in Fig.~\ref{fig:mass_exp}.
(b) Orbital-, site-, and spin-projected self-consistent DFT bands for FeSe in the non-magnetic (NM) and N\'eel-ordered phase for different local moments, including the usual spin-polarized DFT results (CB). Red and blue denote spin-up and spin-down, respectively, and the size of the symbols give the relative weight; all bands are also denoted by gray dots. The projections are for the Fe site with net spin up moment; the other Fe site will be identical, but with the spins reversed. Relative to M, $|k| \le \frac{1}{4} \frac{2\pi}{a} \sim 0.4$~\AA$^{-1}$. Bands for the full zone are given in the Supplemental Material.
}
\label{fig:DFT_bands}
\end{center}
\end{figure}

Here, we reveal a band-theory-based origin for orbital selective physics in monolayer FeSe, which has the highest critical temperature among Fe-based superconductors~\cite{Wang2012Mar,He2013Jul,Zhang2015Jul,Ge2015Mar,Wang2017Mar}. Specifically, monolayer FeSe only has electron pockets near the M point of the Brillouin zone~\cite{Huang2017Mar}, cf.\ Fig.~\ref{fig:DFT_bands}(a). We reveal that the intertwining of the orbital and two-site Fe sublattice degrees of freedom for these electron pockets implies that the electrons in the $x^2-y^2$ orbitals couple strongly to antiferromagnetic (AFM) $Q=(0,0)$ N\'eel, or checkerboard (CB), fluctuations, while those in $\{xz,yz\}$ orbitals do not. This is important in FeSe, where magnetic fluctuations play a central role in superconductivity \cite{Li2016Jun,Agterberg2017Dec,Hirschfeld2011Oct,Chubukov2012Mar,Kreisel2020Aug,Zhai2009Aug,Shishidou2018Mar,Nakayama2018Dec,Huang2017Mar}. In particular, AFM fluctuations with in-plane wavevectors $Q=(\pi,\pi)$ and
$Q=(0,0)$ (in which nearest neighbor Fe-sites have opposite spin) are commonly observed in Fe-based materials.  In
monolayer FeSe, momentum
conservation implies that electrons in the electron pockets can only be coupled by the $Q=(0,0)$ N\'eel fluctuations since there are no hole pockets to allow coupling to the $Q=(\pi,\pi)$ fluctuations.  Consequently, this orbital selective coupling to spin fluctuations provides a natural band-structure-based explanation for the common observation that $x^2-y^2$ electrons are more strongly correlated than $\{xz,yz\}$ electrons in monolayer FeSe \cite{Yi2015Jul,Liu2022Apr}. 

Although the emphasis here is to provide a mechanism for orbital selective physics that relies solely on spin fluctuations, other mechanisms may also play an important role in monolayer FeSe. In particular, the observation of replica bands in previous angle-resolved photoemission spectroscopy (ARPES) measurements has been attributed to the strong coupling of the SrTiO$_3$ (STO) optical phonon modes to the FeSe electrons~\cite{Lee2014Nov,Lee2015Oct,Huang2017Mar,Wang2016Oct}, and we do not consider this coupling here. We also note that while inelastic neutron scattering experiments \cite{Wang2016Jul} observed spin fluctuations in bulk FeSe; similar experiments have not been carried out for monolayer FeSe. Here we argue that the bulk results raise the possibility that spin fluctuations may also be relevant in the two-dimensional material.

In the following, we first present our DFT results that reveal the orbital-selective coupling to magnetic fluctuations. We then present our ARPES data for FeSe$_{1-x}$S$_x$ from which we argue that S doping provides a means to tune the N\'eel fluctuations. We highlight features of the ARPES data for pure monolayer FeSe that are inconsistent with non-magnetic DFT results. Then, after providing the symmetry basis for the orbital selective coupling,  we reveal how orbital selective coupling to N\'eel fluctuations modifies the band structure, leading to better agreement with the ARPES data. Finally, we end with a brief discussion of how our results generalize to other crystal space groups.  

\section{Results and Discussion}
\subsection{Coupling to static Néel order} 
To understand how N\'eel AFM fluctuations couple to the electrons pockets, we performed DFT calculations assuming long-range N\'eel order. These calculations are akin to frozen phonon calculations and allow us to determine the magnitude of the coupling of electronic states to dynamic N\'eel fluctuations.  
In these calculations, shown in Fig.~\ref{fig:DFT_bands}(b) around the M point, the spin-dependent exchange-correlation potential was scaled,
leading to self-consistent local Fe moments ranging between the non-magnetic (NM) and the standard CB result of $\sim$2$\mu_B$. 
As is well-known, the NM DFT bands have a Fermi surface around $\Gamma$, in disagreement with experiments, but for larger moments (see Supplemental Material), the states at $\Gamma$ drop below the Fermi level and there is only the observed electron pocket around M.

Figure~\ref{fig:DFT_bands}(b) provides two key points. The first is that the coupling to the N\'eel order is orbital selective: the band degeneracy of the $x^2-y^2$ orbitals at the M point is split by the N\'eel order, but that of the $\{xz,yz\}$ orbitals is not. The second is that for the $x^2-y^2$ orbitals, this coupling is strong: in the standard CB DFT result, the $x^2-y^2$ bands exhibit a gap of approximately 2~eV and no longer play a role near the chemical potential (and the $\Gamma$ pocket found in the NM calculations -- but not in ARPES experiments -- is removed). At first glance, this orbital selective coupling to N\'eel order is surprising since the N\'eel order is driven by spin-dependent exchange physics and should not reveal a strong orbital dependence. 
This behavior can be intuitively understood, however, by noting that Bloch states having a higher degree of localization on individual Fe sites will be able to couple more strongly to the N\'eel fluctuations. At M, N\'eel interactions, regardless of how weak, cause complete localization of the $x^2-y^2$ states, whereas the degree of localization of the $\{xz,yz\}$ states depends on the size of the local moment.
Additional evidence of the orbital selective behavior is that in the calculations in Fig.~\ref{fig:DFT_bands} for local moments up to $\sim$1 $\mu_B$, the $x^2-y^2$ component of the spin-density matrix is $\sim$50\% larger than the $xz$ or $yz$ component (although the total $\{xz,yz\}$ contribution to the moments is larger).

Since no static, long-range, N\'eel order is observed, we wish to consider dynamic fluctuations of this order. The importance and effects of (symmetry-breaking) fluctuations on the spectral properties -- and the wave functions themselves -- can vary depending on the particular system and property being investigated. For example, it is well-known~\cite{wurtenberg_1990} that to explain the observed broadening of spectral lines in ARPES due to lattice vibrations or the observed exchange splittings above the Curie temperature~\cite{kirschner_1984, kakizaki_1994,aebi_1996, pickel_2010,kagome2024}, it may be necessary to take into account that the local band structure will differ at different places and times in the crystal, i.e., the short-range (broken symmetry) fluctuations. These short-range fluctuations have been treated in different approximations, such as decomposing them into (incoherent) sums of ordered phases, or within an Eliashberg-type formalism in which the single-particle spectra do not explicitly reflect the symmetry-breaking of the fluctuations, an approach that has been used in the interpretations of ARPES spectra for monolayer FeSe \cite{Yi2015Jul,Zhang2016Sep}. 

\begin{figure}[t]
\includegraphics[angle=0,width=1.0\columnwidth]{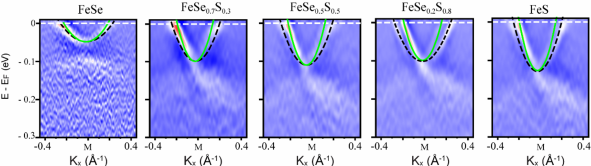}
\caption{Second derivatives of the ARPES intensity around the M point (along the blue line in Fig.~\ref{fig:DFT_bands}(a) and corresponding to the M$\to$$\Gamma$ direction in Fig.~\ref{fig:DFT_bands}(b)) for normal state single-layer FeSe$_{1-x}$S$_x$/STO films.
The extracted band dispersions, indicated by green and dashed black curves, are extracted from four Lorentzian fits of the energy-dependent, momentum-dependent cut curves.
}
\label{fig:mass_exp}
\end{figure}

\subsection{ARPES results}
To gain insight from experiments on the putative role of N\'eel fluctuations, we examined ARPES spectra, Fig.~\ref{fig:mass_exp}, near the M point for single-layer FeSe$_{1-x}$S$_x$/STO films, where the film quality and Se/S concentrations were verified by STM (see Supplemental Material). The alloying of Se and S allows for a systematic assessment of how chemically induced changes in bonding affect the properties, and the ability of models, such as those described below, to account for these variations.  Consideration of the data reveals three noteworthy characteristics. The first is that the effective mass for the bands that cross the Fermi level is largest for FeSe and tends to decrease as the S concentration increases. Secondly, the position of the bands at M relative to the Fermi level, $\mu$, drops as the S concentration increases, but the values of $k_F$ remain approximately constant, consistent with Luttinger's theorem,  which for isoelectronic substitution and a single Fermi surface then locks the band positions relative to $\mu$ to the effective mass. Finally, the spectra around M consist of a set of upper bands dispersing through the Fermi level and lower bands dispersing downward, with no evidence that the two sets of bands cross. For FeSe, there is an apparent gap at M and the lower band (below $\sim$0.1 eV) is almost flat; for FeSe$_{1-x}$S$_x$, the downward dispersion increases and the gap becomes difficult to resolve. The emergence of a gap at the M point for FeSe disagrees with the NM DFT result shown in Fig.~\ref{fig:DFT_bands}(b), which reveals a band crossing of the $x^2-y^2$ and the (flat) $\{xz,yz\}$ bands along M-$\Gamma$. The emergence of this gap has been discussed previously \cite{Yi2015Jul}, where it was pointed out that if the positions of the $\{xz,yz\}$ and the $x^2-y^2$ bands just below the Fermi level are interchanged, then a gap will emerge due to an avoided band crossing. Physically, this was attributed to an orbital selective Mott transition in which electrons in $x^2-y^2$ orbitals become more strongly correlated, flattening this band. These features of the ARPES spectra are also obtained in spin-orbit DFT calculations for the ordered N\'eel phase.

The changes in the experimental positions and effective masses of the bands of FeSe$_{1-x}$S$_x$ reflect differences in standard band structure contributions (e.g., hopping and $p,d$ level positions), N\'eel fluctuations, and other correlation effects. Previous spin-spiral calculations \cite{Shishidou2018Mar} of the Heisenberg parameters place both FeSe and FeS in the region of the phase diagram where quantum fluctuations lead to a disordered (i.e., no long-range order) phase, but because the $J_2/J_1$ value is closer to the critical point for FeSe ($J_2/J_1=0.59$) than FeS ($J_2/J_1=0.69$), N\'eel fluctuations in FeSe are expected to be more important than in FeS~\cite{Wang2015Nov}. This suggests that N\'eel fluctuations play an important role in the strong electronic renormalizations seen in the FeSe spectrum in Fig.~\ref{fig:mass_exp}.

\subsection{Symmetry-required orbital selectivity}
Here, we show that the origin of this orbital selective coupling seen in Fig.~\ref{fig:DFT_bands} lies in the symmetry properties of the Bloch states formed at the M point from $x^2-y^2$ and $\{xz,yz\}$ orbitals.  

FeSe belongs to the non-symmorphic space group 129, $P4/nmm$, which has point group $D_{4h}$, and the Fe atoms occupy Wyckoff position 2a, with site-symmetry $D_{2d}$. At the M point in the NM case, when spin-orbit coupling (SOC) is neglected, there are four different two-dimensional irreducible representations (IRs), labeled $M_1,M_2,M_3,M_4$. Using the notation of the  Bilbao crystallographic server~\cite{Aroyo:2006,Aroyo2:2006}, the $x^2-y^2$ orbitals at the two Fe sites form a $M_1$ IR, while the $\{xz,yz\}$ orbitals contribute to both a $M_3$ and a $M_4$ IR. In the DFT calculations for the non-magnetic state, the $\{xz,yz\}$ states at the M point closest to the chemical potential form a $M_4$ IR, and the corresponding $M_3$ IR appears approximately 1~eV lower. 

To understand whether these electrons at the M point can couple to the N\'eel AFM fluctuations, selection rules can be used. This requires compatibility of the symmetry of the N\'eel AFM state and the symmetry of the bilinear fermion operators that can couple to this state.
The first step is to determine the spatial symmetry of the N\'eel AFM state, which shares the same translation symmetry as the nonmagnetic state and therefore belongs to an IR of the point group $D_{4h}$. The IR must be odd under symmetry operations that exchange the two Fe sites, while even under operations that keep the two sites unchanged. Therefore, the spatial part corresponds to $B_{2u}$ symmetry. The next step is to determine the symmetry of the spin part. In the absence of spin-orbit coupling, the spin symmetry corresponds to $\Gamma_A^S$, which is the axial vector IR of spin-rotations. Therefore, the checkerboard magnetic order transforms as $B_{2u}\otimes \Gamma_A^S$. The symmetry of possible fermion bilinear operators determines the form of the Hamiltonian and how the bilinears couple to the antiferromagnetic order. This is found by taking the direct products of the IRs. Here $M_1\otimes M_1=A_{1g}\oplus B_{2g}\oplus B_{2u}\oplus A_{1u}$, $M_{3}\otimes M_{3}=M_{4}\otimes M_{4}=A_{1g}\oplus B_{2g}\oplus B_{1u}\oplus A_{2u}$, and $M_{3}\otimes M_{4}=A_{2g}\oplus B_{1g}\oplus A_{1u}\oplus B_{2u}$. This reveals that the coupling of the $x^2-y^2$ orbitals to the N\'eel fluctuations is {\it within} the $M_1$ IR (intra-band), as the bilinear contains the $B_{2u}$ IR. However, the coupling of the $\{xz,yz\}$ orbitals to the N\'eel fluctuations has no intra-band component, since the single representations $M_3 \otimes M_3$ and $M_4 \otimes M_4$ do not contain the $B_{2u}$ IR, and therefore the coupling to the Néel AFM fluctuations only occurs {\it between} the $M_3$ and $M_4$ IRs (purely inter-band).

\begin{figure}[t]
\begin{center}
\includegraphics[width=1.0\columnwidth]{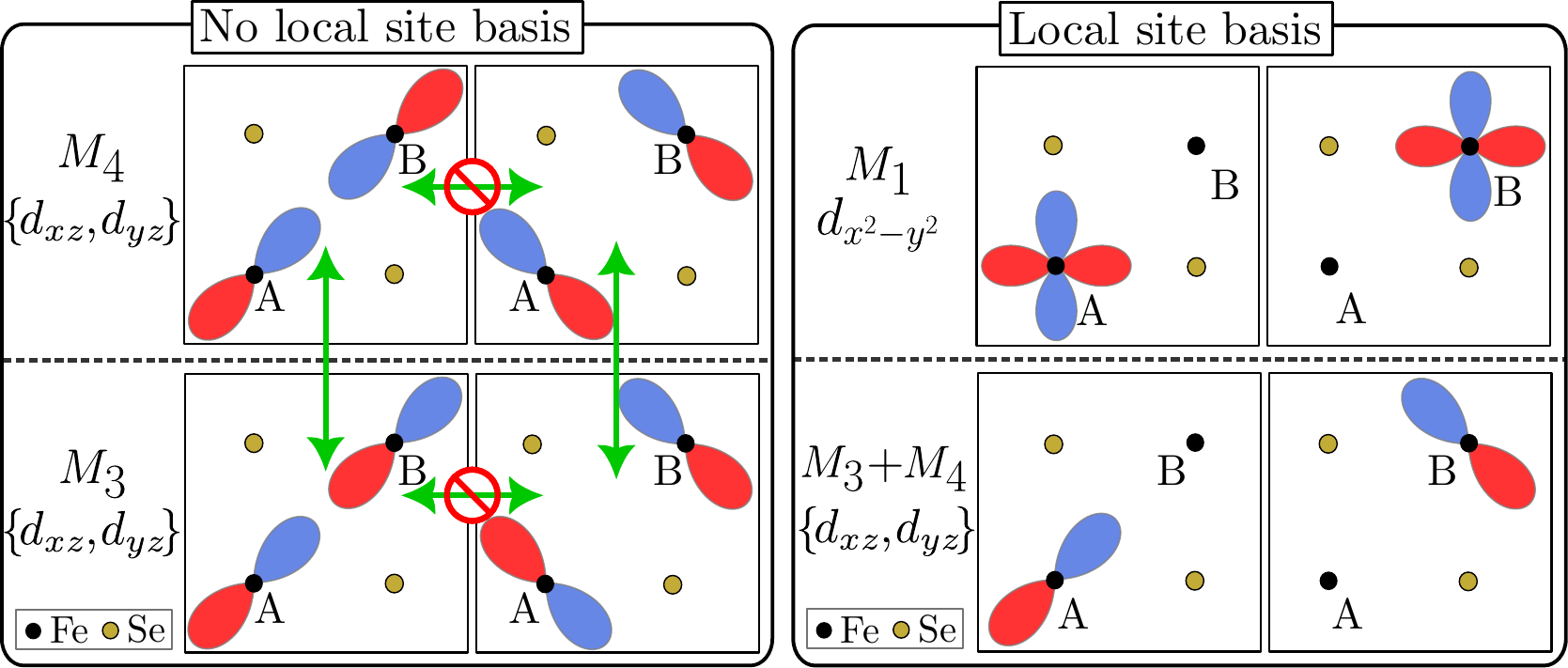} 
\caption{Sketch of the degenerate basis functions for the M point representations $M_1$, $M_3$, $M_4$ and for the hybridization of the $M_3$ and $M_4$ representations. The green arrows show the couplings resulting from the N\'eel ordering.
}
\label{fig:schematic_orbital_selectiviy}
\end{center}
\end{figure}

Equivalently, note that explicitly including the presence of N\'eel order, modifies the symmetry group: the half of the (factor) group operations that relate the two Fe sites are now associated with a spin-flip. As a consequence, the symmetry of the system \textit{in each spin channel} is now the symmorphic space group 115, $P\bar{4}m2$; the two spin channels are related by $PT$  or any of the other spin-flip operations.  Importantly, the Fe site-symmetries (Wyckoff positions 1a and 1b) remain $D_{2d}$. In $D_{2d}$, the $\{xz,yz\}$ orbitals remain in a two-dimensional IR, while the $x^2-y^2$ orbitals (and the other $d$ orbitals) belong to one-dimensional IRs. Thus at M, as seen in Fig.~\ref{fig:DFT_bands}, the $x^2-y^2$ orbitals split, while the $\{xz,yz\}$ orbitals remain degenerate. However, in contrast to the NM case, all the $\{xz,yz\}$ orbitals belong to the same IR.

To gain intuitive insight into this orbital selective coupling, it is useful to examine the Fe site and orbital decomposition of the Bloch wave functions.  Because the $x^2-y^2$ orbitals belong to a 1-D IR of $D_{2d}$, the Hamiltonian around M for these orbitals can be written in a site ($\tau_i$), spin ($\sigma_i$) basis as
\begin{equation}
 H_{x^2-y^2} = \sigma_0 ( \epsilon_{g,\kv} \tau_0^{x^2-y^2} + t_\kv \, \tau_x^{x^2-y^2} ) + B \sigma_z \tau_z^{x^2-y^2} ,
 \label{eq:x2y2}
\end{equation}
where $t_\kv$ is the coupling (hopping) between sites and $B$ is the staggered/N\'eel field. The eigenvalues are $\epsilon_{g,\kv} \pm \sqrt{t_\kv^2 + B^2}$, or at the M point, where $t_\kv\sim k_xk_y$, $\epsilon_{g,\kv} \pm B$. Since $H_{x^2-y^2}$ is diagonal at M, the wave functions are site localized, as illustrated in Fig.~\ref{fig:schematic_orbital_selectiviy}.

For the $\{xz,yz\}$ orbitals, in contrast, the space is inherently 4-dimensional: site and $xz$, $yz$ (8-dimensional including spin). For the NM symmetry at M, this block-diagonalizes into the $M_3$ and $M_4$ IRs, separated by $\sim$1.2 eV. For the $M_4$/$M_3$ IRs, the two degenerate states can be chosen as $|xz+yz,A\rangle \mp|xz+yz,B\rangle,|xz-yz,A\rangle \pm|xz-yz,B\rangle$, upper/lower sign respectively, where $A,B$ are the two Fe sites in the unit cell, cf.\ left panels of Fig.~\ref{fig:schematic_orbital_selectiviy}. It is not possible to make site-localized orbitals from linear combinations of just the two $M_4$ (or $M_3$) degenerate states. In the presence of N\'eel order, however, all the $\{xz,yz\}$ orbitals belong to the same (two-dimensional) IR, and importantly, only specific partners of the $M_4$ and $M_3$ IRs interact as shown in Fig.~\ref{fig:schematic_orbital_selectiviy}. This interband coupling due to the N\'eel ordering both maintains the degeneracies and results in site localization of the eigenfunctions that increases with the strength of the N\'eel coupling; for the CB DFT calculations (with moments $\sim$2$\mu_B$), the site localization of the $\{xz,yz\}$ states is over 90\%.

\subsection{Band renormalization from N\'eel order}
To understand how the AFM N\'eel
fluctuations modify the electronic bands, we provide an effective theory for the electronic states with momenta near the M point. Ignoring spin-orbit coupling, these take the same form, cf. Eq.~\ref{eq:x2y2}, for the two-dimensional $M_1$ ($x^2-y^2$) and $M_4$ ($\{xz,yz\}$) IRs~\cite{Cvetkovic2013Oct,Agterberg2017Dec,Suh2023Sep}:
\begin{equation}
    \begin{aligned}
    H_{i} = & \ \frac{1}{2m_i}(k_x^2+k_y^2)\tau^i_0 - \mu_i - \alpha_i k_x k_y \tau^i_x, 
    \end{aligned}
    \label{eq:kp_M12}
\end{equation}
where $i=\{x^2-y^2,\{xz,yz\}\}$ and the $\tau^i_j$ matrices are Pauli matrices that describe the two-fold degenerate IRs formed from orbitals labeled by $i$ at the M point. The dispersion of both IRs is $\varepsilon_{i,\kv}^\pm = \frac{1}{2m_i}(k_x^2+k_y^2) - \mu_i \pm \alpha_i k_x k_y$.

\begin{figure}
\includegraphics[width=1.0\columnwidth]{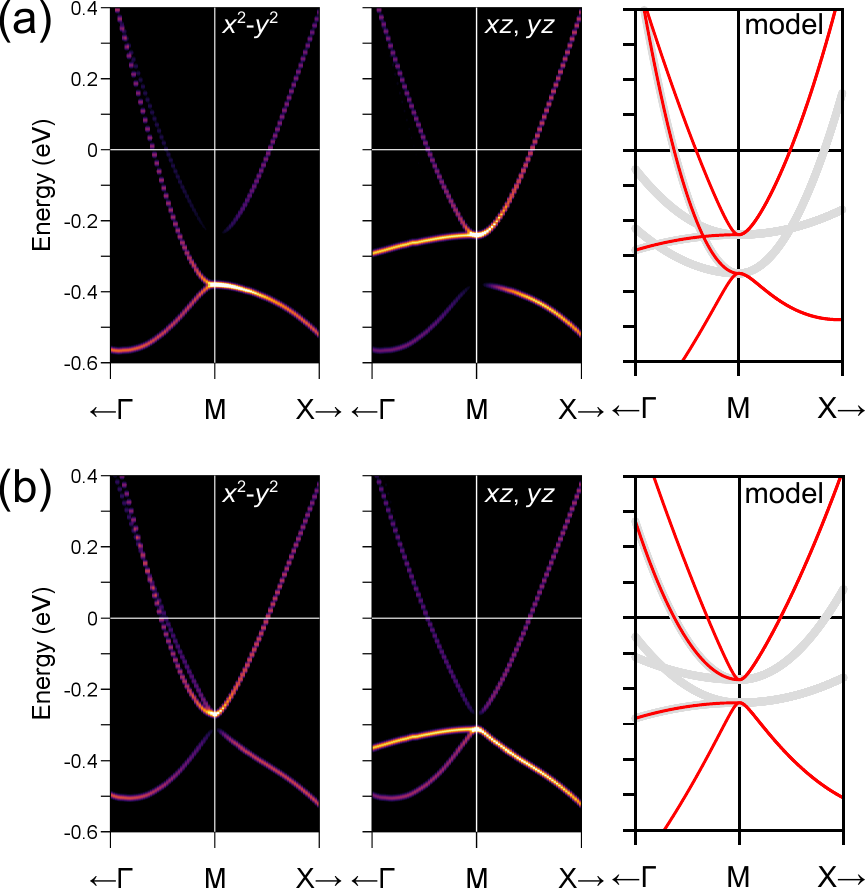}
\caption{Orbital-projected  and $kp$ model bands around 
the M point, $|k| \le \frac{1}{4} \frac{2\pi}{a} \sim 0.4$~\AA$^{-1}$.
(a) Standard NM DFT bands, and model bands using Eq.~\eqref{eq:kp_M12} (third panel) with parameters chosen to approximately fit the NM DFT results:
$m$ = 0.15625 (1.125) (eV-\AA$^2$)$^{-1}$, $\alpha$ = 4.8 (-1.44) eV-\AA$^2$, $\mu$ = 0.35 (0.24) eV for the $x^2-y^2$ ($xz,yz$) orbitals, and a coupling between the $x^2-y^2$ and $\{xz,yz\}$ IRs of $v = 1.12$ eV-\AA, Eq.~\eqref{eq:hybridization_kp}.
(b) Self-consistent bands for a 0.35 eV $x^2-y^2$ self-energy  and model calculations for $Z^{-1}=2$, both values were chosen to produce $\sim$50 meV gaps at M, as observed in the FeSe ARPES data in Fig.~\ref{fig:mass_exp}.  Gray model bands are for $v$=0, showing the importance of the coupling between the two sets of orbitals to remove the crossing between the bands.}
\label{fig:bands_renorm}
\end{figure}

Although uncoupled at the M point, for momenta away from this point these two IRs are coupled through 
\begin{equation}
    H_v = v \Gamma_y (k_x \tau^{\rm mix}_0 + k_y \tau^{\rm mix}_x),
    \label{eq:hybridization_kp}
\end{equation}
where $\Gamma_y$ is a Pauli matrix coupling the two different IRs and $\tau_i^{\rm mix}$ are Pauli matrices operating within the IRs. As shown in Fig.~\ref{fig:bands_renorm}, the NM DFT bands can be approximately reproduced by an appropriate choice of the parameters.  Here, motivated by our symmetry arguments and DFT results in Fig.~\ref{fig:DFT_bands}, we will only include coupling of the N\'eel fluctuations  to the $x^2-y^2$ bands. In addition to generating finite lifetimes for the $x^2-y^2$ quasiparticles, these fluctuations will then renormalize the parameters $m_{x^2-y^2}$, $\mu_{x^2-y^2}$, and $\alpha_{x^2-y^2}$. 

Symmetry dictates that the minimal coupling between the AFM N\'eel fluctuations and the $x^2-y^2$ electronic states is described by
\begin{equation}
    g_1 \sum_{\kv,\qv} \, \vec{S}_{-\qv} \, \cdot\, \Psi^\dagger_{M_1,\kv +\qv/2} \, \tau_z^{x^2-y^2} \, \vec{\sigma} \, \Psi_{M_1,\kv-\qv/2} ,
    \label{eq:CB_x2-y2} 
\end{equation}
where $g_1$ is the interaction strength between the electrons and the magnon spin degrees of freedom $\vec{S}_{\qv}$~\cite{Agterberg2017Dec,Abanov2003Mar}, and the basis for $\Psi^T_{M_1,\kv}=(c_{\kv,A,\uparrow},c_{\kv,A,\downarrow},c_{\kv,B,\uparrow},c_{\kv,B,\downarrow})$ describes $x^2-y^2$ orbitals in the basis given in Fig.~\ref{fig:schematic_orbital_selectiviy}. In the case of the $M_4$ $\{xz,yz\}$ IR, the intraband coupling to the AFM Néel fluctuations corresponds to
\begin{equation}
    g_2 \!\!\sum_{\kv,\qv} (\cos k_x - \cos k_y)\vec{S}_{-\qv} \cdot \Psi^\dagger_{M_4,\kv +\qv/2} \, \tau_z^{xz/yz} \, \vec{\sigma} \, \Psi_{M_4,\kv-\qv/2} ,
    \label{eq:CB_M4} 
\end{equation}
which is suppressed at the M point due to the $(\cos k_x - \cos k_y)$ form factor. However, there is an interband momentum-independent coupling between the $M_3$ and $M_4$ IRs for $\{xz,yz\}$ orbitals,
\begin{equation}
    g_3 \!\!\sum_{\kv,\qv} \vec{S}_{-\qv} \cdot \Psi^\dagger_{M_3,\kv +\qv/2} \, \tau_0^{M_3,M_4} \, \vec{\sigma} \, \Psi_{M_4,\kv-\qv/2} + \textrm{h.c.},
    \label{eq:CB_M4_M3} 
\end{equation}
where $\textrm{h.c.}$ denotes the hermitian conjugate and $\tau_0^{M_3,M_4}$ is the identity matrix in the $4\cross 4$ space including the $M_3$ and $M_4$ IRs. The basis for $\Psi_{M_4,\kv-\qv/2}$ and $ \Psi_{M_3,\kv-\qv/2}$ are detailed in Fig.~\ref{fig:schematic_orbital_selectiviy}.

In the following we take $g_2=g_3=0$ and consider only the coupling of the $x^2-y^2$ orbitals to the N\'eel fluctuations $g_1$ since our DFT calculations indicate this is the dominant coupling. We analyze the first-order, or one-loop, self-energy correction, which has previously been considered a good approximation for cuprates~\cite{Eschrig2000Oct,Vilk1997Nov} and has been used to understand the origin of Fermi surface shrinkage due to electronic interactions observed in Fe-based pnictides~\cite{Ortenzi:2009,Bhattacharyya:2020}. In this case, the coupling to magnetic fluctuations in Eq.~\eqref{eq:CB_x2-y2} anticommutes with the $\tau^{x^2-y^2}_x$ term in the Hamiltonian in Eq.~\eqref{eq:kp_M12}, and therefore this corresponds to an inter-band interacting vertex between the two bands $\varepsilon_{x^2-y^2,\kv}^{+}$ and $\varepsilon_{x^ 2-y^2,\kv}^{-}$. Hence, this yields a self-energy for the fermions in each band,
\begin{equation}
    \Sigma^{\pm}\!(\kv,i\omega_n)\! =\!{-} \frac{g^2}{\beta}\!\! \sum_{\qv,i\nu_n}\! G_0^{\mp}\! (\qv,i\nu_n)  D_0 (\kv-\qv,i\omega_n-i\nu_n),
\end{equation}
where $G_0^\mp(\qv,i\nu_n) = (i\nu_n - \varepsilon_{x^ 2-y^2,\qv}^{\mp})^{-1}$ is the non-interacting fermion propagator. The AFM N\'eel fluctuations are included as magnons with a non-interacting propagator $D_0(\qv,i\nu_n) = \frac{1}{i\nu_n - \Omega_{\qv}} - \frac{1}{i\nu_n + \Omega_{\qv}}$, with $\Omega_{\qv}$ the momentum-dependent frequency of the magnon mode.  

Obtaining realistic parameters to accurately determine the band renormalization is not straightforward for two reasons. First, while the dispersion for the magnon mode has been measured for the tetragonal phase of bulk FeSe~\cite{Wang2016Jul}, it has not been measured in monolayer FeSe. Secondly, there are in principle  additional renormalization effects, as discussed in Refs.~\cite{Ortenzi:2009,Chang2025Jul}. Given these uncertainties, we carry out a simplified approach, allowing only for a magnon mode at a single frequency $\omega_0$ and examining the nature of the renormalizations in this limit. Therefore, the self-energy corrections can be taken to be only frequency dependent.

In this limit, we have two self-energies, $\Sigma_+(i\omega_n)$ and $\Sigma_-(i\omega_n)$. Due to the form of the band dispersion written below Eq.~\eqref{eq:kp_M12}, and the interband nature of the vertex, the two bands are renormalized in the same way, $\Sigma^+(i\omega_n) = \Sigma^-(i\omega_n)$.
In this case, the effective mass $m_{x^2-y^2}^*$  and $\alpha_{x^2-y^2}^*$  are renormalized by the same factor $Z^{-1}= 1 -  \left. \frac{\partial \Re \Sigma (\omega)}{\partial\omega} \right\rvert_{\omega = 0} = N(0) g^2/\omega_0$, with $N(0)$ the density of states at $\mu=0$, $g$ the electron-magnon constant and $\omega_0$ the frequency of the magnon mode. Hence,  $\frac{m_{x^2-y^2}^*}{m_{x^2-y^2}} = Z^{-1}$ and $\frac{\alpha_{x^2-y^2}^*}{\alpha_{x^2-y^2}} = Z$. 

To get an estimate for $Z^{-1}$, we choose $N(0)=0.21$~states/eV per Fe, which reproduces the bands and the Fermi surface for FeSe in Fig.~\ref{fig:mass_exp}, and we estimate $\omega_0=0.05$~eV for the frequency of the magnon mode~\cite{Wang2016Jul}. From our DFT calculations, we identified the electron-magnon constant $g=0.93$~eV. With these parameters, we obtain a renormalization factor $Z^{-1} \approx 3.6$. Although there are uncertainties in $Z^{-1}$, the result we find is consistent with the  difference between the experimental results in Fig.~\ref{fig:mass_exp} and the NM DFT result in Fig~\ref{fig:DFT_bands}.

While our estimate for $Z^{-1}$ for the $x^2-y^2$ orbitals is subject to uncertainty, we expect that our result that the renormalization for the $x^2-y^2$ orbitals due to N\'eel fluctuations is significantly larger than that for the $\{xz,yz\}$ orbitals is robust. For the $\{xz,yz\}$ orbitals, $Z^{-1}$ is strongly suppressed by the form factor in Eq.~\eqref{eq:CB_M4} for intraband N\'eel fluctuations and by the large band energy splitting (1.2 eV) in Eq.~\eqref{eq:CB_M4_M3} for interband N\'eel fluctuations.

Given $Z^{-1}$, the renormalized parameter $\mu_{x^2-y^2}$ will be dictated by Luttinger's theorem. Specifically, the renormalized $m^*_{x^2-y^2}$ and $\alpha^*_{x^2-y^2}$ will push up the $x^2-y^2$ band energy at the M point, enabling a band crossing with the $\{xz,yz\}$ bands. In the rightmost panels of Fig.~\ref{fig:bands_renorm}, we have implemented this approach using the $kp$ theory of Eqs.~\eqref{eq:kp_M12}-\eqref{eq:hybridization_kp}, choosing unrenormalized parameters that approximately reproduce the NM DFT results. We find that choosing $Z^{-1}=2$ to renormalize $m^*_{x^2-y^2}$ and $\alpha^*_{x^2-y^2}$ and  applying the Luttinger argument to renormalize $\mu_{x^2-y^2}$ yields a gapped dispersion, consistent with our ARPES results for FeSe in Fig.~\ref{fig:mass_exp}. (The variations in the bands with S concentration observed in Fig.~\ref{fig:mass_exp} can be mimicked by modifying the model parameters, including decreasing $Z^{-1}$ to account for the expectation mentioned above that the N\'eel fluctuations are less important in FeS.)

Although the above $kp$-theory-based approach is consistent with our experimental results, we note that the renormalization of $\mu_{x^2-y^2}$ can be offset by a renormalization of $\mu_{\{xz,yz\}}$, and we have assumed that there is no such renormalization. To verify the Luttinger mechanism we propose here, in which the renormalizations of $m^*_{x^2-y^2}$ and $\alpha^*_{x^2-y^2}$ drive a renormalization only of $\mu_{x^2-y^2}$ (and not of  $\mu_{\{xz,yz\}}$), we have also used a DFT-based approach (left panels of Fig.~\ref{fig:bands_renorm}).  In particular, we introduce an orbital-dependent real self-energy for the $x^2-y^2$ orbitals and perform self-consistent DFT calculations; in this way, hybridization effects and all band/chemical potential shifts are automatically included. 
As discussed above, for the NM case the $x^2-y^2$ states lie below the $\{xz,yz\}$. As the strength of the orbital-selective interaction increases, the $x^2-y^2$ orbitals rise until a gap is formed. This occurs despite the existence of a $\Gamma$ pocket in the DFT calculations. Hence, all $k_F$s remain essentially constant for the electron pockets, in agreement with the Luttinger's theorem argument we used above. The increased effective mass of the upward dispersing $x^2-y^2$ state along M--$\Gamma$ is evident, as is the increased $x^2-y^2$ weight at the chemical potential.

We note that the renormalization factor $Z^{-1}=2$ used above, while accounting for the  appearance of a gap, does not account for the full mass enhancement observed in the experimental data in Fig.~\ref{fig:mass_exp}, or all the features of the band structure throughout the Brillouin zone. This can be understood as a consequence of additional sources of band renormalization.  Specifically, it has been found that both the $x^2-y^2$ and the $\{xz,yz\}$ bands are renormalized by additional renormalization effects, likely Coulombic in origin, that appear at higher energies~\cite{Ortenzi:2009,Chang2025Jul,Zou2025Jun}.

\subsection{Superconductivity}

N\'eel fluctuations have  been shown to stabilize a $d$-wave superconductor for FeSe~\cite{Li2016Jun,Agterberg2017Dec,Nakayama2018Dec,Huang2017Mar,Ge2019Apr}.  Specifically, these fluctuations give rise to a nodeless $d$-wave state in which the Fermi surface is fully gapped, but the gap changes sign between the two ellipses forming the Fermi surface. Currently, it is believed that the superconducting state in monolayer FeSe is either this nodeless $d$-wave state or a $s$-wave superconducting state \cite{Huang2017Mar}. In the context of a nodeless $d$-wave state, our observation that $x^2-y^2$ electrons are more strongly coupled to N\'eel fluctuations indicate that these electrons play the dominant role in forming a $d$-wave superconductor. The net result of the renormalizations due to the N\'eel fluctuations argued above is to introduce more $x^2-y^2$ weight at the chemical potential -- hence amplifying the pairing from the N\'eel fluctuations and providing an explanation as to why this pairing becomes more energetic than the usual $s_{\pm}$ gap. Such a nodeless $d$-wave gap is  consistent with the earlier association of a decreasing role of N\'eel fluctuations with increasing S concentration and the observation of a gap magnitude that decreases with increasing S concentration~\cite{Zou2025Jun}.

\subsection{Orbital selective coupling in other space groups}
Formally underlying the orbital selective physics described above was that different orbitals belong to different IRs of the little group at the M point. 
Specifically, this difference requires that the $x^2-y^2$ orbitals form a site-local basis while the $\{xz,yz\}$ orbitals do not. It is natural to ask whether this can occur for other crystal space groups. This will be examined in more detail elsewhere, but the answer is yes. We have found there are similar examples, other tetragonal space groups, such as in the rutile structure (space group 136) at the M point or the A point. In addition, we have also found that this applies to hexagonal space groups. For example, in a closed packed hexagonal material (space group 194) at the point H, electrons in different $d$-orbitals will exhibit a related orbital selective coupling to nematic order. 

\section{Conclusions}
We have identified a new band-theory-based mechanism for orbital-selective physics in monolayer FeSe. Our DFT calculations support this mechanism by showing that, for the electron pockets in FeSe, AFM fluctuations couple strongly to $x^2-y^2$ orbitals, but weakly to $\{xz,yz\}$ orbitals. We have revealed that this difference can be understood by examining the symmetry properties of these electronic  Bloch states, and stems from the fact that $x^2-y^2$ orbitals are intrinsically Fe-site localized, allowing the coupling to the N\'eel AFM order. To analyze the consequences of the strong coupling of the $x^2-y^2$ orbitals to the magnetic order, we have estimated band renormalizations from the first-order self-energy correction and show that they can account for the difference between DFT calculations and observed ARPES results. 
Finally, we highlight that the mechanism identified here for symmetry-required orbital selectivity in FeSe can be generalized to other space groups, which will allow new applications for our results on orbital-selective physics.

\section{Methods}
\noindent\textbf{Molecular beam epitaxy growth:} The single-layer FeSe$_{x}$S$_{1-x}$ films were grown on Nb-doped (0.5\% wt) SrTiO$_3$ (001) substrates by molecular beam epitaxy. Fe was supplied via an electron beam source, S and Se via two separate Knudsen cells. The growth rate of films was 0.5 layer/min. The Se(S) concentration is tuned by adjusting the Se(S) cell temperature. The flux ratio of Fe:Se(S) is around 1:10. During growth, the substrates were heated at around 250$^\circ$C, and then the as-grown samples were annealed at around 450$^\circ$C for 1 hour.

\noindent\textbf{Angle-resolved photoemission spectroscopy:}
The ARPES measurements were carried out using a Scienta DA30-L analyzer and He gas discharge lamp ($h\nu = 21.2$~eV) at 80~K. The energy resolution was set at 5~meV and the angular resolution 0.25$^\circ$.
The replica bands seen previously are not resolved here because of temperature effects \cite{Yang2019}.

\subsection{Acknowledgements}
We thank Brian M. Andersen, Rafael Fernandes and Jian Wang for useful discussions.
Work at UWM and WVU was supported by National Science Foundation Grant No.\ DMREF 2323857 and No.\ DMREF 2323858.
M.R. acknowledges support from the Simons Foundation grant SFI-MPS-NFS-00006741-02. D.F.A. and Y. Y. were also supported by the U.S. Department of Energy, Office of Basic Energy Sciences, Division of Materials Sciences and Engineering under Award No. DE-SC0021971 for symmetry-based calculations.

\subsection{Author contributions}
M.R., D.F.A., L.L. and M.W. conceived and designed the study.
M.R., Y.Y., D.F.A and M.W. developed the theoretical model and performed the analytical calculations.
Q.Z. and L.L. performed the ARPES measurements and ARPES data analysis with assistance from B.D.O and H.Z. T.S. and M.W. performed the DFT calculations.
M.R., D.F.A., L.L. and M.W. wrote the paper with input from all authors.

\textit{{\bf Correspondence}} and requests for materials should be addressed to Mercè Roig (roigserv@uwm.edu)

\subsection{Competing interest}
The authors declare no competing interests.

\subsection{Data availability}
The data that support the findings of this study are available from the corresponding
author upon reasonable request.

\subsection{Code availability}
The code that supports the findings of the study is available from the corresponding authors upon reasonable request.

\bibliography{FeSe}

\begin{thebibliography}{50}%
\makeatletter
\providecommand \@ifxundefined [1]{%
 \@ifx{#1\undefined}
}%
\providecommand \@ifnum [1]{%
 \ifnum #1\expandafter \@firstoftwo
 \else \expandafter \@secondoftwo
 \fi
}%
\providecommand \@ifx [1]{%
 \ifx #1\expandafter \@firstoftwo
 \else \expandafter \@secondoftwo
 \fi
}%
\providecommand \natexlab [1]{#1}%
\providecommand \enquote  [1]{``#1''}%
\providecommand \bibnamefont  [1]{#1}%
\providecommand \bibfnamefont [1]{#1}%
\providecommand \citenamefont [1]{#1}%
\providecommand \href@noop [0]{\@secondoftwo}%
\providecommand \href [0]{\begingroup \@sanitize@url \@href}%
\providecommand \@href[1]{\@@startlink{#1}\@@href}%
\providecommand \@@href[1]{\endgroup#1\@@endlink}%
\providecommand \@sanitize@url [0]{\catcode `\\12\catcode `\$12\catcode `\&12\catcode `\#12\catcode `\^12\catcode `\_12\catcode `\%12\relax}%
\providecommand \@@startlink[1]{}%
\providecommand \@@endlink[0]{}%
\providecommand \url  [0]{\begingroup\@sanitize@url \@url }%
\providecommand \@url [1]{\endgroup\@href {#1}{\urlprefix }}%
\providecommand \urlprefix  [0]{URL }%
\providecommand \Eprint [0]{\href }%
\providecommand \doibase [0]{http://dx.doi.org/}%
\providecommand \selectlanguage [0]{\@gobble}%
\providecommand \bibinfo  [0]{\@secondoftwo}%
\providecommand \bibfield  [0]{\@secondoftwo}%
\providecommand \translation [1]{[#1]}%
\providecommand \BibitemOpen [0]{}%
\providecommand \bibitemStop [0]{}%
\providecommand \bibitemNoStop [0]{.\EOS\space}%
\providecommand \EOS [0]{\spacefactor3000\relax}%
\providecommand \BibitemShut  [1]{\csname bibitem#1\endcsname}%
\let\auto@bib@innerbib\@empty
\bibitem [{\citenamefont {Sprau}\ \emph {et~al.}(2017)\citenamefont {Sprau}, \citenamefont {Kostin}, \citenamefont {Kreisel}, \citenamefont {B{\ifmmode\ddot{o}\else\"{o}\fi}hmer}, \citenamefont {Taufour}, \citenamefont {Canfield}, \citenamefont {Mukherjee}, \citenamefont {Hirschfeld}, \citenamefont {Andersen},\ and\ \citenamefont {Davis}}]{Sprau2017Jul}%
  \BibitemOpen
  \bibfield  {author} {\bibinfo {author} {\bibfnamefont {P.~O.}\ \bibnamefont {Sprau}}, \bibinfo {author} {\bibfnamefont {A.}~\bibnamefont {Kostin}}, \bibinfo {author} {\bibfnamefont {A.}~\bibnamefont {Kreisel}}, \bibinfo {author} {\bibfnamefont {A.~E.}\ \bibnamefont {B{\ifmmode\ddot{o}\else\"{o}\fi}hmer}}, \bibinfo {author} {\bibfnamefont {V.}~\bibnamefont {Taufour}}, \bibinfo {author} {\bibfnamefont {P.~C.}\ \bibnamefont {Canfield}}, \bibinfo {author} {\bibfnamefont {S.}~\bibnamefont {Mukherjee}}, \bibinfo {author} {\bibfnamefont {P.~J.}\ \bibnamefont {Hirschfeld}}, \bibinfo {author} {\bibfnamefont {B.~M.}\ \bibnamefont {Andersen}}, \ and\ \bibinfo {author} {\bibfnamefont {J.~C.~S{\ifmmode\acute{e}\else\'{e}\fi}amus}\ \bibnamefont {Davis}},\ }\bibfield  {title} {\enquote {\bibinfo {title} {{Discovery of orbital-selective Cooper pairing in FeSe}},}\ }\href {\doibase 10.1126/science.aal1575} {\bibfield  {journal} {\bibinfo  {journal} {Science}\ }\textbf {\bibinfo {volume} {357}},\ \bibinfo {pages} {75}
  (\bibinfo {year} {2017})}\BibitemShut {NoStop}%
\bibitem [{\citenamefont {Kreisel}\ \emph {et~al.}(2017)\citenamefont {Kreisel}, \citenamefont {Andersen}, \citenamefont {Sprau}, \citenamefont {Kostin}, \citenamefont {Davis},\ and\ \citenamefont {Hirschfeld}}]{Kreisel2017May}%
  \BibitemOpen
  \bibfield  {author} {\bibinfo {author} {\bibfnamefont {Andreas}\ \bibnamefont {Kreisel}}, \bibinfo {author} {\bibfnamefont {Brian~M.}\ \bibnamefont {Andersen}}, \bibinfo {author} {\bibfnamefont {P.~O.}\ \bibnamefont {Sprau}}, \bibinfo {author} {\bibfnamefont {A.}~\bibnamefont {Kostin}}, \bibinfo {author} {\bibfnamefont {J.~C.~S\'eamus}\ \bibnamefont {Davis}}, \ and\ \bibinfo {author} {\bibfnamefont {P.~J.}\ \bibnamefont {Hirschfeld}},\ }\bibfield  {title} {\enquote {\bibinfo {title} {{Orbital selective pairing and gap structures of iron-based superconductors}},}\ }\href {\doibase 10.1103/PhysRevB.95.174504} {\bibfield  {journal} {\bibinfo  {journal} {Phys. Rev. B}\ }\textbf {\bibinfo {volume} {95}},\ \bibinfo {pages} {174504} (\bibinfo {year} {2017})}\BibitemShut {NoStop}%
\bibitem [{\citenamefont {Liu}\ \emph {et~al.}(2022)\citenamefont {Liu}, \citenamefont {Kreisel}, \citenamefont {Zhong}, \citenamefont {Li}, \citenamefont {Andersen}, \citenamefont {Hirschfeld},\ and\ \citenamefont {Wang}}]{Liu2022Apr}%
  \BibitemOpen
  \bibfield  {author} {\bibinfo {author} {\bibfnamefont {Chaofei}\ \bibnamefont {Liu}}, \bibinfo {author} {\bibfnamefont {Andreas}\ \bibnamefont {Kreisel}}, \bibinfo {author} {\bibfnamefont {Shan}\ \bibnamefont {Zhong}}, \bibinfo {author} {\bibfnamefont {Yu}~\bibnamefont {Li}}, \bibinfo {author} {\bibfnamefont {Brian~M.}\ \bibnamefont {Andersen}}, \bibinfo {author} {\bibfnamefont {Peter}\ \bibnamefont {Hirschfeld}}, \ and\ \bibinfo {author} {\bibfnamefont {Jian}\ \bibnamefont {Wang}},\ }\bibfield  {title} {\enquote {\bibinfo {title} {{Orbital-Selective High-Temperature Cooper Pairing Developed in the Two-Dimensional Limit}},}\ }\href {\doibase 10.1021/acs.nanolett.1c04863} {\bibfield  {journal} {\bibinfo  {journal} {Nano Lett.}\ }\textbf {\bibinfo {volume} {22}},\ \bibinfo {pages} {3245} (\bibinfo {year} {2022})}\BibitemShut {NoStop}%
\bibitem [{\citenamefont {Kostin}\ \emph {et~al.}(2018)\citenamefont {Kostin}, \citenamefont {Sprau}, \citenamefont {Kreisel}, \citenamefont {Chong}, \citenamefont {B{\ifmmode\ddot{o}\else\"{o}\fi}hmer}, \citenamefont {Canfield}, \citenamefont {Hirschfeld}, \citenamefont {Andersen},\ and\ \citenamefont {Davis}}]{Kostin2018Oct}%
  \BibitemOpen
  \bibfield  {author} {\bibinfo {author} {\bibfnamefont {A.}~\bibnamefont {Kostin}}, \bibinfo {author} {\bibfnamefont {P.~O.}\ \bibnamefont {Sprau}}, \bibinfo {author} {\bibfnamefont {A.}~\bibnamefont {Kreisel}}, \bibinfo {author} {\bibfnamefont {Yi~Xue}\ \bibnamefont {Chong}}, \bibinfo {author} {\bibfnamefont {A.~E.}\ \bibnamefont {B{\ifmmode\ddot{o}\else\"{o}\fi}hmer}}, \bibinfo {author} {\bibfnamefont {P.~C.}\ \bibnamefont {Canfield}}, \bibinfo {author} {\bibfnamefont {P.~J.}\ \bibnamefont {Hirschfeld}}, \bibinfo {author} {\bibfnamefont {B.~M.}\ \bibnamefont {Andersen}}, \ and\ \bibinfo {author} {\bibfnamefont {J.~C.~S{\ifmmode\acute{e}\else\'{e}\fi}amus}\ \bibnamefont {Davis}},\ }\bibfield  {title} {\enquote {\bibinfo {title} {{Imaging orbital-selective quasiparticles in the Hund{'}s metal state of FeSe}},}\ }\href {\doibase 10.1038/s41563-018-0151-0} {\bibfield  {journal} {\bibinfo  {journal} {Nat. Mater.}\ }\textbf {\bibinfo {volume} {17}},\ \bibinfo {pages} {869} (\bibinfo {year} {2018})}\BibitemShut
  {NoStop}%
\bibitem [{\citenamefont {Yi}\ \emph {et~al.}(2015)\citenamefont {Yi}, \citenamefont {Liu}, \citenamefont {Zhang}, \citenamefont {Yu}, \citenamefont {Zhu}, \citenamefont {Lee}, \citenamefont {Moore}, \citenamefont {Schmitt}, \citenamefont {Li}, \citenamefont {Riggs}, \citenamefont {Chu}, \citenamefont {Lv}, \citenamefont {Hu}, \citenamefont {Hashimoto}, \citenamefont {Mo}, \citenamefont {Hussain}, \citenamefont {Mao}, \citenamefont {Chu}, \citenamefont {Fisher}, \citenamefont {Si}, \citenamefont {Shen},\ and\ \citenamefont {Lu}}]{Yi2015Jul}%
  \BibitemOpen
  \bibfield  {author} {\bibinfo {author} {\bibfnamefont {M.}~\bibnamefont {Yi}}, \bibinfo {author} {\bibfnamefont {Z.-K.}\ \bibnamefont {Liu}}, \bibinfo {author} {\bibfnamefont {Y.}~\bibnamefont {Zhang}}, \bibinfo {author} {\bibfnamefont {R.}~\bibnamefont {Yu}}, \bibinfo {author} {\bibfnamefont {J.-X.}\ \bibnamefont {Zhu}}, \bibinfo {author} {\bibfnamefont {J.~J.}\ \bibnamefont {Lee}}, \bibinfo {author} {\bibfnamefont {R.~G.}\ \bibnamefont {Moore}}, \bibinfo {author} {\bibfnamefont {F.~T.}\ \bibnamefont {Schmitt}}, \bibinfo {author} {\bibfnamefont {W.}~\bibnamefont {Li}}, \bibinfo {author} {\bibfnamefont {S.~C.}\ \bibnamefont {Riggs}}, \bibinfo {author} {\bibfnamefont {J.-H.}\ \bibnamefont {Chu}}, \bibinfo {author} {\bibfnamefont {B.}~\bibnamefont {Lv}}, \bibinfo {author} {\bibfnamefont {J.}~\bibnamefont {Hu}}, \bibinfo {author} {\bibfnamefont {M.}~\bibnamefont {Hashimoto}}, \bibinfo {author} {\bibfnamefont {S.-K.}\ \bibnamefont {Mo}}, \bibinfo {author} {\bibfnamefont {Z.}~\bibnamefont {Hussain}}, \bibinfo
  {author} {\bibfnamefont {Z.~Q.}\ \bibnamefont {Mao}}, \bibinfo {author} {\bibfnamefont {C.~W.}\ \bibnamefont {Chu}}, \bibinfo {author} {\bibfnamefont {I.~R.}\ \bibnamefont {Fisher}}, \bibinfo {author} {\bibfnamefont {Q.}~\bibnamefont {Si}}, \bibinfo {author} {\bibfnamefont {Z.-X.}\ \bibnamefont {Shen}}, \ and\ \bibinfo {author} {\bibfnamefont {D.~H.}\ \bibnamefont {Lu}},\ }\bibfield  {title} {\enquote {\bibinfo {title} {{Observation of universal strong orbital-dependent correlation effects in iron chalcogenides}},}\ }\href {\doibase 10.1038/ncomms8777} {\bibfield  {journal} {\bibinfo  {journal} {Nat. Commun.}\ }\textbf {\bibinfo {volume} {6}},\ \bibinfo {pages} {7777} (\bibinfo {year} {2015})}\BibitemShut {NoStop}%
\bibitem [{\citenamefont {De{'}~Medici}\ \emph {et~al.}(2014)\citenamefont {De{'}~Medici}, \citenamefont {Giovannetti},\ and\ \citenamefont {Capone}}]{De'Medici2014Apr}%
  \BibitemOpen
  \bibfield  {author} {\bibinfo {author} {\bibfnamefont {Luca}\ \bibnamefont {De{'}~Medici}}, \bibinfo {author} {\bibfnamefont {Gianluca}\ \bibnamefont {Giovannetti}}, \ and\ \bibinfo {author} {\bibfnamefont {Massimo}\ \bibnamefont {Capone}},\ }\bibfield  {title} {\enquote {\bibinfo {title} {{Selective Mott Physics as a Key to Iron Superconductors}},}\ }\href {\doibase 10.1103/PhysRevLett.112.177001} {\bibfield  {journal} {\bibinfo  {journal} {Phys. Rev. Lett.}\ }\textbf {\bibinfo {volume} {112}},\ \bibinfo {pages} {177001} (\bibinfo {year} {2014})}\BibitemShut {NoStop}%
\bibitem [{\citenamefont {Huang}\ \emph {et~al.}(2022)\citenamefont {Huang}, \citenamefont {Yu}, \citenamefont {Xu}, \citenamefont {Zhu}, \citenamefont {Oh}, \citenamefont {Jiang}, \citenamefont {Wang}, \citenamefont {Wu}, \citenamefont {Chen}, \citenamefont {Denlinger}, \citenamefont {Mo}, \citenamefont {Hashimoto}, \citenamefont {Michiardi}, \citenamefont {Pedersen}, \citenamefont {Gorovikov}, \citenamefont {Zhdanovich}, \citenamefont {Damascelli}, \citenamefont {Gu}, \citenamefont {Dai}, \citenamefont {Chu}, \citenamefont {Lu}, \citenamefont {Si}, \citenamefont {Birgeneau},\ and\ \citenamefont {Yi}}]{Huang2022Jan}%
  \BibitemOpen
  \bibfield  {author} {\bibinfo {author} {\bibfnamefont {Jianwei}\ \bibnamefont {Huang}}, \bibinfo {author} {\bibfnamefont {Rong}\ \bibnamefont {Yu}}, \bibinfo {author} {\bibfnamefont {Zhijun}\ \bibnamefont {Xu}}, \bibinfo {author} {\bibfnamefont {Jian-Xin}\ \bibnamefont {Zhu}}, \bibinfo {author} {\bibfnamefont {Ji~Seop}\ \bibnamefont {Oh}}, \bibinfo {author} {\bibfnamefont {Qianni}\ \bibnamefont {Jiang}}, \bibinfo {author} {\bibfnamefont {Meng}\ \bibnamefont {Wang}}, \bibinfo {author} {\bibfnamefont {Han}\ \bibnamefont {Wu}}, \bibinfo {author} {\bibfnamefont {Tong}\ \bibnamefont {Chen}}, \bibinfo {author} {\bibfnamefont {Jonathan~D.}\ \bibnamefont {Denlinger}}, \bibinfo {author} {\bibfnamefont {Sung-Kwan}\ \bibnamefont {Mo}}, \bibinfo {author} {\bibfnamefont {Makoto}\ \bibnamefont {Hashimoto}}, \bibinfo {author} {\bibfnamefont {Matteo}\ \bibnamefont {Michiardi}}, \bibinfo {author} {\bibfnamefont {Tor~M.}\ \bibnamefont {Pedersen}}, \bibinfo {author} {\bibfnamefont {Sergey}\ \bibnamefont {Gorovikov}}, \bibinfo
  {author} {\bibfnamefont {Sergey}\ \bibnamefont {Zhdanovich}}, \bibinfo {author} {\bibfnamefont {Andrea}\ \bibnamefont {Damascelli}}, \bibinfo {author} {\bibfnamefont {Genda}\ \bibnamefont {Gu}}, \bibinfo {author} {\bibfnamefont {Pengcheng}\ \bibnamefont {Dai}}, \bibinfo {author} {\bibfnamefont {Jiun-Haw}\ \bibnamefont {Chu}}, \bibinfo {author} {\bibfnamefont {Donghui}\ \bibnamefont {Lu}}, \bibinfo {author} {\bibfnamefont {Qimiao}\ \bibnamefont {Si}}, \bibinfo {author} {\bibfnamefont {Robert~J.}\ \bibnamefont {Birgeneau}}, \ and\ \bibinfo {author} {\bibfnamefont {Ming}\ \bibnamefont {Yi}},\ }\bibfield  {title} {\enquote {\bibinfo {title} {{Correlation-driven electronic reconstruction in FeTe1{-}xSex}},}\ }\href {\doibase 10.1038/s42005-022-00805-6} {\bibfield  {journal} {\bibinfo  {journal} {Commun. Phys.}\ }\textbf {\bibinfo {volume} {5}},\ \bibinfo {pages} {1--9} (\bibinfo {year} {2022})}\BibitemShut {NoStop}%
\bibitem [{\citenamefont {Yu}\ \emph {et~al.}(2021)\citenamefont {Yu}, \citenamefont {Hu}, \citenamefont {Nica}, \citenamefont {Zhu},\ and\ \citenamefont {Si}}]{fphys21}%
  \BibitemOpen
  \bibfield  {author} {\bibinfo {author} {\bibfnamefont {Rong}\ \bibnamefont {Yu}}, \bibinfo {author} {\bibfnamefont {Haoyu}\ \bibnamefont {Hu}}, \bibinfo {author} {\bibfnamefont {Emilian~M.}\ \bibnamefont {Nica}}, \bibinfo {author} {\bibfnamefont {Jian-Xin}\ \bibnamefont {Zhu}}, \ and\ \bibinfo {author} {\bibfnamefont {Qimiao}\ \bibnamefont {Si}},\ }\bibfield  {title} {\enquote {\bibinfo {title} {Orbital selectivity in electron correlations and superconducting pairing of iron-based superconductors},}\ }\href {\doibase 10.3389/fphy.2021.578347} {\bibfield  {journal} {\bibinfo  {journal} {Frontiers in Physics}\ }\textbf {\bibinfo {volume} {9}},\ \bibinfo {pages} {578347} (\bibinfo {year} {2021})}\BibitemShut {NoStop}%
\bibitem [{\citenamefont {Kiesel}\ and\ \citenamefont {Thomale}(2012)}]{Kiesel:2012}%
  \BibitemOpen
  \bibfield  {author} {\bibinfo {author} {\bibfnamefont {Maximilian~L.}\ \bibnamefont {Kiesel}}\ and\ \bibinfo {author} {\bibfnamefont {Ronny}\ \bibnamefont {Thomale}},\ }\bibfield  {title} {\enquote {\bibinfo {title} {Sublattice interference in the kagome hubbard model},}\ }\href {\doibase 10.1103/PhysRevB.86.121105} {\bibfield  {journal} {\bibinfo  {journal} {Phys. Rev. B}\ }\textbf {\bibinfo {volume} {86}},\ \bibinfo {pages} {121105} (\bibinfo {year} {2012})}\BibitemShut {NoStop}%
\bibitem [{\citenamefont {Kiesel}\ \emph {et~al.}(2013)\citenamefont {Kiesel}, \citenamefont {Platt},\ and\ \citenamefont {Thomale}}]{Kiesel:2013}%
  \BibitemOpen
  \bibfield  {author} {\bibinfo {author} {\bibfnamefont {Maximilian~L.}\ \bibnamefont {Kiesel}}, \bibinfo {author} {\bibfnamefont {Christian}\ \bibnamefont {Platt}}, \ and\ \bibinfo {author} {\bibfnamefont {Ronny}\ \bibnamefont {Thomale}},\ }\bibfield  {title} {\enquote {\bibinfo {title} {Unconventional fermi surface instabilities in the kagome hubbard model},}\ }\href {\doibase 10.1103/PhysRevLett.110.126405} {\bibfield  {journal} {\bibinfo  {journal} {Phys. Rev. Lett.}\ }\textbf {\bibinfo {volume} {110}},\ \bibinfo {pages} {126405} (\bibinfo {year} {2013})}\BibitemShut {NoStop}%
\bibitem [{\citenamefont {Mielke}\ \emph {et~al.}(2022)\citenamefont {Mielke}, \citenamefont {Das}, \citenamefont {Yin}, \citenamefont {Liu}, \citenamefont {Gupta}, \citenamefont {Jiang}, \citenamefont {Medarde}, \citenamefont {Wu}, \citenamefont {Lei}, \citenamefont {Chang}, \citenamefont {Dai}, \citenamefont {Si}, \citenamefont {Miao}, \citenamefont {Thomale}, \citenamefont {Neupert}, \citenamefont {Shi}, \citenamefont {Khasanov}, \citenamefont {Hasan}, \citenamefont {Luetkens},\ and\ \citenamefont {Guguchia}}]{Mielke:2022}%
  \BibitemOpen
  \bibfield  {author} {\bibinfo {author} {\bibfnamefont {C.}~\bibnamefont {Mielke}}, \bibinfo {author} {\bibfnamefont {D.}~\bibnamefont {Das}}, \bibinfo {author} {\bibfnamefont {J.-X.}\ \bibnamefont {Yin}}, \bibinfo {author} {\bibfnamefont {H.}~\bibnamefont {Liu}}, \bibinfo {author} {\bibfnamefont {R.}~\bibnamefont {Gupta}}, \bibinfo {author} {\bibfnamefont {Y.-X.}\ \bibnamefont {Jiang}}, \bibinfo {author} {\bibfnamefont {M.}~\bibnamefont {Medarde}}, \bibinfo {author} {\bibfnamefont {X.}~\bibnamefont {Wu}}, \bibinfo {author} {\bibfnamefont {H.~C.}\ \bibnamefont {Lei}}, \bibinfo {author} {\bibfnamefont {J.}~\bibnamefont {Chang}}, \bibinfo {author} {\bibfnamefont {Pengcheng}\ \bibnamefont {Dai}}, \bibinfo {author} {\bibfnamefont {Q.}~\bibnamefont {Si}}, \bibinfo {author} {\bibfnamefont {H.}~\bibnamefont {Miao}}, \bibinfo {author} {\bibfnamefont {R.}~\bibnamefont {Thomale}}, \bibinfo {author} {\bibfnamefont {T.}~\bibnamefont {Neupert}}, \bibinfo {author} {\bibfnamefont {Y.}~\bibnamefont {Shi}}, \bibinfo {author}
  {\bibfnamefont {R.}~\bibnamefont {Khasanov}}, \bibinfo {author} {\bibfnamefont {M.~Z.}\ \bibnamefont {Hasan}}, \bibinfo {author} {\bibfnamefont {H.}~\bibnamefont {Luetkens}}, \ and\ \bibinfo {author} {\bibfnamefont {Z.}~\bibnamefont {Guguchia}},\ }\bibfield  {title} {\enquote {\bibinfo {title} {Time-reversal symmetry-breaking charge order in a kagome superconductor},}\ }\href {\doibase 10.1038/s41586-021-04327-z} {\bibfield  {journal} {\bibinfo  {journal} {Nature}\ }\textbf {\bibinfo {volume} {602}},\ \bibinfo {pages} {245–250} (\bibinfo {year} {2022})}\BibitemShut {NoStop}%
\bibitem [{\citenamefont {{Q.-Y. Wang, Z. Li, W.-H. Zhang, Z.-C. Zhang, J.-S. Zhang, W. Li, H. Ding, Y.-B. Ou, P. Deng, K. Chang, J. Wen, C.-L. Song, K. He, J.-F. Jia, S.-H. Ji, Y.-Y. Wang, L.-L. Wang, X. Chen, X.-C. Ma, and Q.-K. Xue}}(2012)}]{Wang2012Mar}%
  \BibitemOpen
  \bibfield  {author} {\bibinfo {author} {\bibnamefont {{Q.-Y. Wang, Z. Li, W.-H. Zhang, Z.-C. Zhang, J.-S. Zhang, W. Li, H. Ding, Y.-B. Ou, P. Deng, K. Chang, J. Wen, C.-L. Song, K. He, J.-F. Jia, S.-H. Ji, Y.-Y. Wang, L.-L. Wang, X. Chen, X.-C. Ma, and Q.-K. Xue}}},\ }\bibfield  {title} {\enquote {\bibinfo {title} {{Interface-Induced High-Temperature Superconductivity in Single Unit-Cell FeSe Films on SrTiO$_3$}},}\ }\href {\doibase 10.1088/0256-307X/29/3/037402} {\bibfield  {journal} {\bibinfo  {journal} {Chin. Phys. Lett.}\ }\textbf {\bibinfo {volume} {29}},\ \bibinfo {pages} {037402} (\bibinfo {year} {2012})}\BibitemShut {NoStop}%
\bibitem [{\citenamefont {He}\ \emph {et~al.}(2013)\citenamefont {He}, \citenamefont {He}, \citenamefont {Zhang}, \citenamefont {Zhao}, \citenamefont {Liu}, \citenamefont {Liu}, \citenamefont {Mou}, \citenamefont {Ou}, \citenamefont {Wang}, \citenamefont {Li}, \citenamefont {Wang}, \citenamefont {Peng}, \citenamefont {Liu}, \citenamefont {Chen}, \citenamefont {Yu}, \citenamefont {Liu}, \citenamefont {Dong}, \citenamefont {Zhang}, \citenamefont {Chen}, \citenamefont {Xu}, \citenamefont {Chen}, \citenamefont {Ma}, \citenamefont {Xue},\ and\ \citenamefont {Zhou}}]{He2013Jul}%
  \BibitemOpen
  \bibfield  {author} {\bibinfo {author} {\bibfnamefont {Shaolong}\ \bibnamefont {He}}, \bibinfo {author} {\bibfnamefont {Junfeng}\ \bibnamefont {He}}, \bibinfo {author} {\bibfnamefont {Wenhao}\ \bibnamefont {Zhang}}, \bibinfo {author} {\bibfnamefont {Lin}\ \bibnamefont {Zhao}}, \bibinfo {author} {\bibfnamefont {Defa}\ \bibnamefont {Liu}}, \bibinfo {author} {\bibfnamefont {Xu}~\bibnamefont {Liu}}, \bibinfo {author} {\bibfnamefont {Daixiang}\ \bibnamefont {Mou}}, \bibinfo {author} {\bibfnamefont {Yun-Bo}\ \bibnamefont {Ou}}, \bibinfo {author} {\bibfnamefont {Qing-Yan}\ \bibnamefont {Wang}}, \bibinfo {author} {\bibfnamefont {Zhi}\ \bibnamefont {Li}}, \bibinfo {author} {\bibfnamefont {Lili}\ \bibnamefont {Wang}}, \bibinfo {author} {\bibfnamefont {Yingying}\ \bibnamefont {Peng}}, \bibinfo {author} {\bibfnamefont {Yan}\ \bibnamefont {Liu}}, \bibinfo {author} {\bibfnamefont {Chaoyu}\ \bibnamefont {Chen}}, \bibinfo {author} {\bibfnamefont {Li}~\bibnamefont {Yu}}, \bibinfo {author} {\bibfnamefont {Guodong}\
  \bibnamefont {Liu}}, \bibinfo {author} {\bibfnamefont {Xiaoli}\ \bibnamefont {Dong}}, \bibinfo {author} {\bibfnamefont {Jun}\ \bibnamefont {Zhang}}, \bibinfo {author} {\bibfnamefont {Chuangtian}\ \bibnamefont {Chen}}, \bibinfo {author} {\bibfnamefont {Zuyan}\ \bibnamefont {Xu}}, \bibinfo {author} {\bibfnamefont {Xi}~\bibnamefont {Chen}}, \bibinfo {author} {\bibfnamefont {Xucun}\ \bibnamefont {Ma}}, \bibinfo {author} {\bibfnamefont {Qikun}\ \bibnamefont {Xue}}, \ and\ \bibinfo {author} {\bibfnamefont {X.~J.}\ \bibnamefont {Zhou}},\ }\bibfield  {title} {\enquote {\bibinfo {title} {{Phase diagram and electronic indication of high-temperature superconductivity at 65{\hspace{0.167em}}K in single-layer FeSe films}},}\ }\href {\doibase 10.1038/nmat3648} {\bibfield  {journal} {\bibinfo  {journal} {Nat. Mater.}\ }\textbf {\bibinfo {volume} {12}},\ \bibinfo {pages} {605} (\bibinfo {year} {2013})}\BibitemShut {NoStop}%
\bibitem [{\citenamefont {Zhang}\ \emph {et~al.}(2015)\citenamefont {Zhang}, \citenamefont {Wang}, \citenamefont {Song}, \citenamefont {Liu}, \citenamefont {Peng}, \citenamefont {Moler}, \citenamefont {Feng},\ and\ \citenamefont {Wang}}]{Zhang2015Jul}%
  \BibitemOpen
  \bibfield  {author} {\bibinfo {author} {\bibfnamefont {Zuocheng}\ \bibnamefont {Zhang}}, \bibinfo {author} {\bibfnamefont {Yi-Hua}\ \bibnamefont {Wang}}, \bibinfo {author} {\bibfnamefont {Qi}~\bibnamefont {Song}}, \bibinfo {author} {\bibfnamefont {Chang}\ \bibnamefont {Liu}}, \bibinfo {author} {\bibfnamefont {Rui}\ \bibnamefont {Peng}}, \bibinfo {author} {\bibfnamefont {K.~A.}\ \bibnamefont {Moler}}, \bibinfo {author} {\bibfnamefont {Donglai}\ \bibnamefont {Feng}}, \ and\ \bibinfo {author} {\bibfnamefont {Yayu}\ \bibnamefont {Wang}},\ }\bibfield  {title} {\enquote {\bibinfo {title} {{Onset of the Meissner effect at 65K in FeSe thin film grown on Nb-doped SrTiO$_3$ substrate}},}\ }\href {\doibase 10.1007/s11434-015-0842-8} {\bibfield  {journal} {\bibinfo  {journal} {Science Bulletin}\ }\textbf {\bibinfo {volume} {60}},\ \bibinfo {pages} {1301} (\bibinfo {year} {2015})}\BibitemShut {NoStop}%
\bibitem [{\citenamefont {Ge}\ \emph {et~al.}(2015)\citenamefont {Ge}, \citenamefont {Liu}, \citenamefont {Liu}, \citenamefont {Gao}, \citenamefont {Qian}, \citenamefont {Xue}, \citenamefont {Liu},\ and\ \citenamefont {Jia}}]{Ge2015Mar}%
  \BibitemOpen
  \bibfield  {author} {\bibinfo {author} {\bibfnamefont {Jian-Feng}\ \bibnamefont {Ge}}, \bibinfo {author} {\bibfnamefont {Zhi-Long}\ \bibnamefont {Liu}}, \bibinfo {author} {\bibfnamefont {Canhua}\ \bibnamefont {Liu}}, \bibinfo {author} {\bibfnamefont {Chun-Lei}\ \bibnamefont {Gao}}, \bibinfo {author} {\bibfnamefont {Dong}\ \bibnamefont {Qian}}, \bibinfo {author} {\bibfnamefont {Qi-Kun}\ \bibnamefont {Xue}}, \bibinfo {author} {\bibfnamefont {Ying}\ \bibnamefont {Liu}}, \ and\ \bibinfo {author} {\bibfnamefont {Jin-Feng}\ \bibnamefont {Jia}},\ }\bibfield  {title} {\enquote {\bibinfo {title} {{Superconductivity above 100 K in single-layer FeSe films on doped SrTiO$_3$}},}\ }\href {\doibase 10.1038/nmat4153} {\bibfield  {journal} {\bibinfo  {journal} {Nat. Mater.}\ }\textbf {\bibinfo {volume} {14}},\ \bibinfo {pages} {285} (\bibinfo {year} {2015})}\BibitemShut {NoStop}%
\bibitem [{\citenamefont {Wang}\ \emph {et~al.}(2017)\citenamefont {Wang}, \citenamefont {Liu}, \citenamefont {Liu},\ and\ \citenamefont {Wang}}]{Wang2017Mar}%
  \BibitemOpen
  \bibfield  {author} {\bibinfo {author} {\bibfnamefont {Ziqiao}\ \bibnamefont {Wang}}, \bibinfo {author} {\bibfnamefont {Chaofei}\ \bibnamefont {Liu}}, \bibinfo {author} {\bibfnamefont {Yi}~\bibnamefont {Liu}}, \ and\ \bibinfo {author} {\bibfnamefont {Jian}\ \bibnamefont {Wang}},\ }\bibfield  {title} {\enquote {\bibinfo {title} {{High-temperature superconductivity in one-unit-cell FeSe films}},}\ }\href {\doibase 10.1088/1361-648X/aa5f26} {\bibfield  {journal} {\bibinfo  {journal} {J. Phys.: Condens. Matter}\ }\textbf {\bibinfo {volume} {29}},\ \bibinfo {pages} {153001} (\bibinfo {year} {2017})}\BibitemShut {NoStop}%
\bibitem [{\citenamefont {Huang}\ and\ \citenamefont {Hoffman}(2017)}]{Huang2017Mar}%
  \BibitemOpen
  \bibfield  {author} {\bibinfo {author} {\bibfnamefont {Dennis}\ \bibnamefont {Huang}}\ and\ \bibinfo {author} {\bibfnamefont {Jennifer~E}\ \bibnamefont {Hoffman}},\ }\bibfield  {title} {\enquote {\bibinfo {title} {{Monolayer FeSe on SrTiO$_3$}},}\ }\href {\doibase 10.1146/annurev-conmatphys-031016-025242} {\bibfield  {journal} {\bibinfo  {journal} {Annu. Rev. Condens. Matter Phys.}\ }\textbf {\bibinfo {volume} {8}},\ \bibinfo {pages} {311} (\bibinfo {year} {2017})}\BibitemShut {NoStop}%
\bibitem [{\citenamefont {Li}\ \emph {et~al.}(2016)\citenamefont {Li}, \citenamefont {Wang}, \citenamefont {Yao},\ and\ \citenamefont {Lee}}]{Li2016Jun}%
  \BibitemOpen
  \bibfield  {author} {\bibinfo {author} {\bibfnamefont {Zi-Xiang}\ \bibnamefont {Li}}, \bibinfo {author} {\bibfnamefont {Fa}~\bibnamefont {Wang}}, \bibinfo {author} {\bibfnamefont {Hong}\ \bibnamefont {Yao}}, \ and\ \bibinfo {author} {\bibfnamefont {Dung-Hai}\ \bibnamefont {Lee}},\ }\bibfield  {title} {\enquote {\bibinfo {title} {{What makes the T$_c$ of monolayer FeSe on SrTiO$_3$ so high: A sign-problem-free quantum Monte Carlo study}},}\ }\href {\doibase 10.1007/s11434-016-1087-x} {\bibfield  {journal} {\bibinfo  {journal} {Science Bulletin}\ }\textbf {\bibinfo {volume} {61}},\ \bibinfo {pages} {925} (\bibinfo {year} {2016})}\BibitemShut {NoStop}%
\bibitem [{\citenamefont {Agterberg}\ \emph {et~al.}(2017)\citenamefont {Agterberg}, \citenamefont {Shishidou}, \citenamefont {O{'}Halloran}, \citenamefont {Brydon},\ and\ \citenamefont {Weinert}}]{Agterberg2017Dec}%
  \BibitemOpen
  \bibfield  {author} {\bibinfo {author} {\bibfnamefont {D.~F.}\ \bibnamefont {Agterberg}}, \bibinfo {author} {\bibfnamefont {T.}~\bibnamefont {Shishidou}}, \bibinfo {author} {\bibfnamefont {J.}~\bibnamefont {O{'}Halloran}}, \bibinfo {author} {\bibfnamefont {P.~M.~R.}\ \bibnamefont {Brydon}}, \ and\ \bibinfo {author} {\bibfnamefont {M.}~\bibnamefont {Weinert}},\ }\bibfield  {title} {\enquote {\bibinfo {title} {{Resilient Nodeless $d$-Wave Superconductivity in Monolayer FeSe}},}\ }\href {\doibase 10.1103/PhysRevLett.119.267001} {\bibfield  {journal} {\bibinfo  {journal} {Phys. Rev. Lett.}\ }\textbf {\bibinfo {volume} {119}},\ \bibinfo {pages} {267001} (\bibinfo {year} {2017})}\BibitemShut {NoStop}%
\bibitem [{\citenamefont {Hirschfeld}\ \emph {et~al.}(2011)\citenamefont {Hirschfeld}, \citenamefont {Korshunov},\ and\ \citenamefont {Mazin}}]{Hirschfeld2011Oct}%
  \BibitemOpen
  \bibfield  {author} {\bibinfo {author} {\bibfnamefont {P.~J.}\ \bibnamefont {Hirschfeld}}, \bibinfo {author} {\bibfnamefont {M.~M.}\ \bibnamefont {Korshunov}}, \ and\ \bibinfo {author} {\bibfnamefont {I.~I.}\ \bibnamefont {Mazin}},\ }\bibfield  {title} {\enquote {\bibinfo {title} {{Gap symmetry and structure of Fe-based superconductors}},}\ }\href {\doibase 10.1088/0034-4885/74/12/124508} {\bibfield  {journal} {\bibinfo  {journal} {Rep. Prog. Phys.}\ }\textbf {\bibinfo {volume} {74}},\ \bibinfo {pages} {124508} (\bibinfo {year} {2011})}\BibitemShut {NoStop}%
\bibitem [{\citenamefont {Chubukov}(2012)}]{Chubukov2012Mar}%
  \BibitemOpen
  \bibfield  {author} {\bibinfo {author} {\bibfnamefont {Andrey}\ \bibnamefont {Chubukov}},\ }\bibfield  {title} {\enquote {\bibinfo {title} {{Pairing Mechanism in Fe-Based Superconductors}},}\ }\href {\doibase 10.1146/annurev-conmatphys-020911-125055} {\bibfield  {journal} {\bibinfo  {journal} {Annu. Rev. Condens. Matter Phys.}\ }\textbf {\bibinfo {volume} {3}},\ \bibinfo {pages} {57} (\bibinfo {year} {2012})}\BibitemShut {NoStop}%
\bibitem [{\citenamefont {Kreisel}\ \emph {et~al.}(2020)\citenamefont {Kreisel}, \citenamefont {Hirschfeld},\ and\ \citenamefont {Andersen}}]{Kreisel2020Aug}%
  \BibitemOpen
  \bibfield  {author} {\bibinfo {author} {\bibfnamefont {Andreas}\ \bibnamefont {Kreisel}}, \bibinfo {author} {\bibfnamefont {Peter~J.}\ \bibnamefont {Hirschfeld}}, \ and\ \bibinfo {author} {\bibfnamefont {Brian~M.}\ \bibnamefont {Andersen}},\ }\bibfield  {title} {\enquote {\bibinfo {title} {{On the Remarkable Superconductivity of FeSe and Its Close Cousins}},}\ }\href {\doibase 10.3390/sym12091402} {\bibfield  {journal} {\bibinfo  {journal} {Symmetry}\ }\textbf {\bibinfo {volume} {12}},\ \bibinfo {pages} {1402} (\bibinfo {year} {2020})}\BibitemShut {NoStop}%
\bibitem [{\citenamefont {Zhai}\ \emph {et~al.}(2009)\citenamefont {Zhai}, \citenamefont {Wang},\ and\ \citenamefont {Lee}}]{Zhai2009Aug}%
  \BibitemOpen
  \bibfield  {author} {\bibinfo {author} {\bibfnamefont {Hui}\ \bibnamefont {Zhai}}, \bibinfo {author} {\bibfnamefont {Fa}~\bibnamefont {Wang}}, \ and\ \bibinfo {author} {\bibfnamefont {Dung-Hai}\ \bibnamefont {Lee}},\ }\bibfield  {title} {\enquote {\bibinfo {title} {{Antiferromagnetically driven electronic correlations in iron pnictides and cuprates}},}\ }\href {\doibase 10.1103/PhysRevB.80.064517} {\bibfield  {journal} {\bibinfo  {journal} {Phys. Rev. B}\ }\textbf {\bibinfo {volume} {80}},\ \bibinfo {pages} {064517} (\bibinfo {year} {2009})}\BibitemShut {NoStop}%
\bibitem [{\citenamefont {Shishidou}\ \emph {et~al.}(2018)\citenamefont {Shishidou}, \citenamefont {Agterberg},\ and\ \citenamefont {Weinert}}]{Shishidou2018Mar}%
  \BibitemOpen
  \bibfield  {author} {\bibinfo {author} {\bibfnamefont {T.}~\bibnamefont {Shishidou}}, \bibinfo {author} {\bibfnamefont {D.~F.}\ \bibnamefont {Agterberg}}, \ and\ \bibinfo {author} {\bibfnamefont {M.}~\bibnamefont {Weinert}},\ }\bibfield  {title} {\enquote {\bibinfo {title} {{Magnetic fluctuations in single-layer FeSe}},}\ }\href {\doibase 10.1038/s42005-018-0006-7} {\bibfield  {journal} {\bibinfo  {journal} {Commun. Phys.}\ }\textbf {\bibinfo {volume} {1}},\ \bibinfo {pages} {8} (\bibinfo {year} {2018})}\BibitemShut {NoStop}%
\bibitem [{\citenamefont {Nakayama}\ \emph {et~al.}(2018)\citenamefont {Nakayama}, \citenamefont {Shishidou},\ and\ \citenamefont {Agterberg}}]{Nakayama2018Dec}%
  \BibitemOpen
  \bibfield  {author} {\bibinfo {author} {\bibfnamefont {Takeru}\ \bibnamefont {Nakayama}}, \bibinfo {author} {\bibfnamefont {Tatsuya}\ \bibnamefont {Shishidou}}, \ and\ \bibinfo {author} {\bibfnamefont {Daniel~F.}\ \bibnamefont {Agterberg}},\ }\bibfield  {title} {\enquote {\bibinfo {title} {{Nodal topology in $d$-wave superconducting monolayer FeSe}},}\ }\href {\doibase 10.1103/PhysRevB.98.214503} {\bibfield  {journal} {\bibinfo  {journal} {Phys. Rev. B}\ }\textbf {\bibinfo {volume} {98}},\ \bibinfo {pages} {214503} (\bibinfo {year} {2018})}\BibitemShut {NoStop}%
\bibitem [{\citenamefont {Lee}\ \emph {et~al.}(2014)\citenamefont {Lee}, \citenamefont {Schmitt}, \citenamefont {Moore}, \citenamefont {Johnston}, \citenamefont {Cui}, \citenamefont {Li}, \citenamefont {Yi}, \citenamefont {Liu}, \citenamefont {Hashimoto}, \citenamefont {Zhang}, \citenamefont {Lu}, \citenamefont {Devereaux}, \citenamefont {Lee},\ and\ \citenamefont {Shen}}]{Lee2014Nov}%
  \BibitemOpen
  \bibfield  {author} {\bibinfo {author} {\bibfnamefont {J.~J.}\ \bibnamefont {Lee}}, \bibinfo {author} {\bibfnamefont {F.~T.}\ \bibnamefont {Schmitt}}, \bibinfo {author} {\bibfnamefont {R.~G.}\ \bibnamefont {Moore}}, \bibinfo {author} {\bibfnamefont {S.}~\bibnamefont {Johnston}}, \bibinfo {author} {\bibfnamefont {Y.-T.}\ \bibnamefont {Cui}}, \bibinfo {author} {\bibfnamefont {W.}~\bibnamefont {Li}}, \bibinfo {author} {\bibfnamefont {M.}~\bibnamefont {Yi}}, \bibinfo {author} {\bibfnamefont {Z.~K.}\ \bibnamefont {Liu}}, \bibinfo {author} {\bibfnamefont {M.}~\bibnamefont {Hashimoto}}, \bibinfo {author} {\bibfnamefont {Y.}~\bibnamefont {Zhang}}, \bibinfo {author} {\bibfnamefont {D.~H.}\ \bibnamefont {Lu}}, \bibinfo {author} {\bibfnamefont {T.~P.}\ \bibnamefont {Devereaux}}, \bibinfo {author} {\bibfnamefont {D.-H.}\ \bibnamefont {Lee}}, \ and\ \bibinfo {author} {\bibfnamefont {Z.-X.}\ \bibnamefont {Shen}},\ }\bibfield  {title} {\enquote {\bibinfo {title} {{Interfacial mode coupling as the origin of the enhancement
  of Tc in FeSe films on SrTiO3}},}\ }\href {\doibase 10.1038/nature13894} {\bibfield  {journal} {\bibinfo  {journal} {Nature}\ }\textbf {\bibinfo {volume} {515}},\ \bibinfo {pages} {245--248} (\bibinfo {year} {2014})}\BibitemShut {NoStop}%
\bibitem [{\citenamefont {Lee}(2015)}]{Lee2015Oct}%
  \BibitemOpen
  \bibfield  {author} {\bibinfo {author} {\bibfnamefont {Dung-Hai}\ \bibnamefont {Lee}},\ }\bibfield  {title} {\enquote {\bibinfo {title} {{What makes the Tc of FeSe/SrTiO3 so high?}}}\ }\href {\doibase 10.1088/1674-1056/24/11/117405} {\bibfield  {journal} {\bibinfo  {journal} {Chin. Phys. B}\ }\textbf {\bibinfo {volume} {24}},\ \bibinfo {pages} {117405} (\bibinfo {year} {2015})}\BibitemShut {NoStop}%
\bibitem [{\citenamefont {Wang}\ \emph {et~al.}(2016{\natexlab{a}})\citenamefont {Wang}, \citenamefont {Ma},\ and\ \citenamefont {Xue}}]{Wang2016Oct}%
  \BibitemOpen
  \bibfield  {author} {\bibinfo {author} {\bibfnamefont {Lili}\ \bibnamefont {Wang}}, \bibinfo {author} {\bibfnamefont {Xucun}\ \bibnamefont {Ma}}, \ and\ \bibinfo {author} {\bibfnamefont {Qi-Kun}\ \bibnamefont {Xue}},\ }\bibfield  {title} {\enquote {\bibinfo {title} {{Interface high-temperature superconductivity}},}\ }\href {\doibase 10.1088/0953-2048/29/12/123001} {\bibfield  {journal} {\bibinfo  {journal} {Supercond. Sci. Technol.}\ }\textbf {\bibinfo {volume} {29}},\ \bibinfo {pages} {123001} (\bibinfo {year} {2016}{\natexlab{a}})}\BibitemShut {NoStop}%
\bibitem [{\citenamefont {Wang}\ \emph {et~al.}(2016{\natexlab{b}})\citenamefont {Wang}, \citenamefont {Shen}, \citenamefont {Pan}, \citenamefont {Zhang}, \citenamefont {Ikeuchi}, \citenamefont {Iida}, \citenamefont {Christianson}, \citenamefont {Walker}, \citenamefont {Adroja}, \citenamefont {Abdel-Hafiez}, \citenamefont {Chen}, \citenamefont {Chareev}, \citenamefont {Vasiliev},\ and\ \citenamefont {Zhao}}]{Wang2016Jul}%
  \BibitemOpen
  \bibfield  {author} {\bibinfo {author} {\bibfnamefont {Qisi}\ \bibnamefont {Wang}}, \bibinfo {author} {\bibfnamefont {Yao}\ \bibnamefont {Shen}}, \bibinfo {author} {\bibfnamefont {Bingying}\ \bibnamefont {Pan}}, \bibinfo {author} {\bibfnamefont {Xiaowen}\ \bibnamefont {Zhang}}, \bibinfo {author} {\bibfnamefont {K.}~\bibnamefont {Ikeuchi}}, \bibinfo {author} {\bibfnamefont {K.}~\bibnamefont {Iida}}, \bibinfo {author} {\bibfnamefont {A.~D.}\ \bibnamefont {Christianson}}, \bibinfo {author} {\bibfnamefont {H.~C.}\ \bibnamefont {Walker}}, \bibinfo {author} {\bibfnamefont {D.~T.}\ \bibnamefont {Adroja}}, \bibinfo {author} {\bibfnamefont {M.}~\bibnamefont {Abdel-Hafiez}}, \bibinfo {author} {\bibfnamefont {Xiaojia}\ \bibnamefont {Chen}}, \bibinfo {author} {\bibfnamefont {D.~A.}\ \bibnamefont {Chareev}}, \bibinfo {author} {\bibfnamefont {A.~N.}\ \bibnamefont {Vasiliev}}, \ and\ \bibinfo {author} {\bibfnamefont {Jun}\ \bibnamefont {Zhao}},\ }\bibfield  {title} {\enquote {\bibinfo {title} {{Magnetic ground state of
  FeSe}},}\ }\href {\doibase 10.1038/ncomms12182} {\bibfield  {journal} {\bibinfo  {journal} {Nat. Commun.}\ }\textbf {\bibinfo {volume} {7}},\ \bibinfo {pages} {12182} (\bibinfo {year} {2016}{\natexlab{b}})}\BibitemShut {NoStop}%
\bibitem [{\citenamefont {W\"urtenberg}\ \emph {et~al.}(1990)\citenamefont {W\"urtenberg}, \citenamefont {Kuch}, \citenamefont {Langsdorf}, \citenamefont {Dietz},\ and\ \citenamefont {Gerhardt}}]{wurtenberg_1990}%
  \BibitemOpen
  \bibfield  {author} {\bibinfo {author} {\bibfnamefont {J}~\bibnamefont {W\"urtenberg}}, \bibinfo {author} {\bibfnamefont {W}~\bibnamefont {Kuch}}, \bibinfo {author} {\bibfnamefont {S}~\bibnamefont {Langsdorf}}, \bibinfo {author} {\bibfnamefont {E}~\bibnamefont {Dietz}}, \ and\ \bibinfo {author} {\bibfnamefont {U}~\bibnamefont {Gerhardt}},\ }\bibfield  {title} {\enquote {\bibinfo {title} {Modification of photoelectron spectra by lattice and spin disorder},}\ }\href {\doibase 10.1088/0031-8949/41/4/057} {\bibfield  {journal} {\bibinfo  {journal} {Physica Scripta}\ }\textbf {\bibinfo {volume} {41}},\ \bibinfo {pages} {634} (\bibinfo {year} {1990})}\BibitemShut {NoStop}%
\bibitem [{\citenamefont {Kirschner}\ \emph {et~al.}(1984)\citenamefont {Kirschner}, \citenamefont {Gl\"obl}, \citenamefont {Dose},\ and\ \citenamefont {Scheidt}}]{kirschner_1984}%
  \BibitemOpen
  \bibfield  {author} {\bibinfo {author} {\bibfnamefont {J.}~\bibnamefont {Kirschner}}, \bibinfo {author} {\bibfnamefont {M.}~\bibnamefont {Gl\"obl}}, \bibinfo {author} {\bibfnamefont {V.}~\bibnamefont {Dose}}, \ and\ \bibinfo {author} {\bibfnamefont {H.}~\bibnamefont {Scheidt}},\ }\bibfield  {title} {\enquote {\bibinfo {title} {Wave-vector-dependent temperature behavior of empty bands in ferromagnetic iron},}\ }\href {\doibase 10.1103/PhysRevLett.53.612} {\bibfield  {journal} {\bibinfo  {journal} {Phys. Rev. Lett.}\ }\textbf {\bibinfo {volume} {53}},\ \bibinfo {pages} {612} (\bibinfo {year} {1984})}\BibitemShut {NoStop}%
\bibitem [{\citenamefont {Kakizaki}\ \emph {et~al.}(1994)\citenamefont {Kakizaki}, \citenamefont {Fujii}, \citenamefont {Shimada}, \citenamefont {Kamata}, \citenamefont {Ono}, \citenamefont {Park}, \citenamefont {Kinoshita}, \citenamefont {Ishii},\ and\ \citenamefont {Fukutani}}]{kakizaki_1994}%
  \BibitemOpen
  \bibfield  {author} {\bibinfo {author} {\bibfnamefont {A.}~\bibnamefont {Kakizaki}}, \bibinfo {author} {\bibfnamefont {J.}~\bibnamefont {Fujii}}, \bibinfo {author} {\bibfnamefont {K.}~\bibnamefont {Shimada}}, \bibinfo {author} {\bibfnamefont {A.}~\bibnamefont {Kamata}}, \bibinfo {author} {\bibfnamefont {K.}~\bibnamefont {Ono}}, \bibinfo {author} {\bibfnamefont {K.-H.}\ \bibnamefont {Park}}, \bibinfo {author} {\bibfnamefont {T.}~\bibnamefont {Kinoshita}}, \bibinfo {author} {\bibfnamefont {T.}~\bibnamefont {Ishii}}, \ and\ \bibinfo {author} {\bibfnamefont {H.}~\bibnamefont {Fukutani}},\ }\bibfield  {title} {\enquote {\bibinfo {title} {Fluctuating local magnetic moments in ferromagnetic {N}i observed by the spin-resolved resonant photoemission},}\ }\href {\doibase 10.1103/PhysRevLett.72.2781} {\bibfield  {journal} {\bibinfo  {journal} {Phys. Rev. Lett.}\ }\textbf {\bibinfo {volume} {72}},\ \bibinfo {pages} {2781} (\bibinfo {year} {1994})}\BibitemShut {NoStop}%
\bibitem [{\citenamefont {Aebi}\ \emph {et~al.}(1996)\citenamefont {Aebi}, \citenamefont {Kreutz}, \citenamefont {Osterwalder}, \citenamefont {Fasel}, \citenamefont {Schwaller},\ and\ \citenamefont {Schlapbach}}]{aebi_1996}%
  \BibitemOpen
  \bibfield  {author} {\bibinfo {author} {\bibfnamefont {P.}~\bibnamefont {Aebi}}, \bibinfo {author} {\bibfnamefont {T.~J.}\ \bibnamefont {Kreutz}}, \bibinfo {author} {\bibfnamefont {J.}~\bibnamefont {Osterwalder}}, \bibinfo {author} {\bibfnamefont {R.}~\bibnamefont {Fasel}}, \bibinfo {author} {\bibfnamefont {P.}~\bibnamefont {Schwaller}}, \ and\ \bibinfo {author} {\bibfnamefont {L.}~\bibnamefont {Schlapbach}},\ }\bibfield  {title} {\enquote {\bibinfo {title} {k-space mapping of majority and minority bands on the {F}ermi surface of nickel below and above the {C}urie temperature},}\ }\href {\doibase 10.1103/PhysRevLett.76.1150} {\bibfield  {journal} {\bibinfo  {journal} {Phys. Rev. Lett.}\ }\textbf {\bibinfo {volume} {76}},\ \bibinfo {pages} {1150} (\bibinfo {year} {1996})}\BibitemShut {NoStop}%
\bibitem [{\citenamefont {Pickel}\ \emph {et~al.}(2010)\citenamefont {Pickel}, \citenamefont {Schmidt}, \citenamefont {Weinelt},\ and\ \citenamefont {Donath}}]{pickel_2010}%
  \BibitemOpen
  \bibfield  {author} {\bibinfo {author} {\bibfnamefont {M.}~\bibnamefont {Pickel}}, \bibinfo {author} {\bibfnamefont {A.~B.}\ \bibnamefont {Schmidt}}, \bibinfo {author} {\bibfnamefont {M.}~\bibnamefont {Weinelt}}, \ and\ \bibinfo {author} {\bibfnamefont {M.}~\bibnamefont {Donath}},\ }\bibfield  {title} {\enquote {\bibinfo {title} {Magnetic exchange splitting in {F}e above the {C}urie temperature},}\ }\href {\doibase 10.1103/PhysRevLett.104.237204} {\bibfield  {journal} {\bibinfo  {journal} {Phys. Rev. Lett.}\ }\textbf {\bibinfo {volume} {104}},\ \bibinfo {pages} {237204} (\bibinfo {year} {2010})}\BibitemShut {NoStop}%
\bibitem [{\citenamefont {Ren}\ \emph {et~al.}(2024)\citenamefont {Ren}, \citenamefont {Huang}, \citenamefont {Tan}, \citenamefont {Biswas}, \citenamefont {Pulkkinen}, \citenamefont {Zhang}, \citenamefont {Xie}, \citenamefont {Yue}, \citenamefont {Chen}, \citenamefont {Xie}, \citenamefont {Allen}, \citenamefont {Wu}, \citenamefont {Ren}, \citenamefont {Rajapitamahuni}, \citenamefont {Kundu}, \citenamefont {Vescovo}, \citenamefont {Kono}, \citenamefont {Morosan}, \citenamefont {Dai}, \citenamefont {Zhu}, \citenamefont {Si}, \citenamefont {Minár}, \citenamefont {Yan},\ and\ \citenamefont {Yi}}]{kagome2024}%
  \BibitemOpen
  \bibfield  {author} {\bibinfo {author} {\bibfnamefont {Zheng}\ \bibnamefont {Ren}}, \bibinfo {author} {\bibfnamefont {Jianwei}\ \bibnamefont {Huang}}, \bibinfo {author} {\bibfnamefont {Hengxin}\ \bibnamefont {Tan}}, \bibinfo {author} {\bibfnamefont {Ananya}\ \bibnamefont {Biswas}}, \bibinfo {author} {\bibfnamefont {Aki}\ \bibnamefont {Pulkkinen}}, \bibinfo {author} {\bibfnamefont {Yichen}\ \bibnamefont {Zhang}}, \bibinfo {author} {\bibfnamefont {Yaofeng}\ \bibnamefont {Xie}}, \bibinfo {author} {\bibfnamefont {Ziqin}\ \bibnamefont {Yue}}, \bibinfo {author} {\bibfnamefont {Lei}\ \bibnamefont {Chen}}, \bibinfo {author} {\bibfnamefont {Fang}\ \bibnamefont {Xie}}, \bibinfo {author} {\bibfnamefont {Kevin}\ \bibnamefont {Allen}}, \bibinfo {author} {\bibfnamefont {Han}\ \bibnamefont {Wu}}, \bibinfo {author} {\bibfnamefont {Qirui}\ \bibnamefont {Ren}}, \bibinfo {author} {\bibfnamefont {Anil}\ \bibnamefont {Rajapitamahuni}}, \bibinfo {author} {\bibfnamefont {Asish~K.}\ \bibnamefont {Kundu}}, \bibinfo {author}
  {\bibfnamefont {Elio}\ \bibnamefont {Vescovo}}, \bibinfo {author} {\bibfnamefont {Junichiro}\ \bibnamefont {Kono}}, \bibinfo {author} {\bibfnamefont {Emilia}\ \bibnamefont {Morosan}}, \bibinfo {author} {\bibfnamefont {Pengcheng}\ \bibnamefont {Dai}}, \bibinfo {author} {\bibfnamefont {Jian-Xin}\ \bibnamefont {Zhu}}, \bibinfo {author} {\bibfnamefont {Qimiao}\ \bibnamefont {Si}}, \bibinfo {author} {\bibfnamefont {Ján}\ \bibnamefont {Minár}}, \bibinfo {author} {\bibfnamefont {Binghai}\ \bibnamefont {Yan}}, \ and\ \bibinfo {author} {\bibfnamefont {Ming}\ \bibnamefont {Yi}},\ }\bibfield  {title} {\enquote {\bibinfo {title} {Persistent flat band splitting and strong selective band renormalization in a kagome magnet thin film},}\ }\href {\doibase 10.1038/s41467-024-53722-3} {\bibfield  {journal} {\bibinfo  {journal} {Nat. Commun.}\ }\textbf {\bibinfo {volume} {15}},\ \bibinfo {pages} {9376} (\bibinfo {year} {2024})}\BibitemShut {NoStop}%
\bibitem [{\citenamefont {Zhang}\ \emph {et~al.}(2016)\citenamefont {Zhang}, \citenamefont {Lee}, \citenamefont {Moore}, \citenamefont {Li}, \citenamefont {Yi}, \citenamefont {Hashimoto}, \citenamefont {Lu}, \citenamefont {Devereaux}, \citenamefont {Lee},\ and\ \citenamefont {Shen}}]{Zhang2016Sep}%
  \BibitemOpen
  \bibfield  {author} {\bibinfo {author} {\bibfnamefont {Y.}~\bibnamefont {Zhang}}, \bibinfo {author} {\bibfnamefont {J.~J.}\ \bibnamefont {Lee}}, \bibinfo {author} {\bibfnamefont {R.~G.}\ \bibnamefont {Moore}}, \bibinfo {author} {\bibfnamefont {W.}~\bibnamefont {Li}}, \bibinfo {author} {\bibfnamefont {M.}~\bibnamefont {Yi}}, \bibinfo {author} {\bibfnamefont {M.}~\bibnamefont {Hashimoto}}, \bibinfo {author} {\bibfnamefont {D.~H.}\ \bibnamefont {Lu}}, \bibinfo {author} {\bibfnamefont {T.~P.}\ \bibnamefont {Devereaux}}, \bibinfo {author} {\bibfnamefont {D.-H.}\ \bibnamefont {Lee}}, \ and\ \bibinfo {author} {\bibfnamefont {Z.-X.}\ \bibnamefont {Shen}},\ }\bibfield  {title} {\enquote {\bibinfo {title} {{Superconducting Gap Anisotropy in Monolayer FeSe Thin Film}},}\ }\href {\doibase 10.1103/PhysRevLett.117.117001} {\bibfield  {journal} {\bibinfo  {journal} {Phys. Rev. Lett.}\ }\textbf {\bibinfo {volume} {117}},\ \bibinfo {pages} {117001} (\bibinfo {year} {2016})}\BibitemShut {NoStop}%
\bibitem [{\citenamefont {Wang}\ \emph {et~al.}(2015)\citenamefont {Wang}, \citenamefont {Kivelson},\ and\ \citenamefont {Lee}}]{Wang2015Nov}%
  \BibitemOpen
  \bibfield  {author} {\bibinfo {author} {\bibfnamefont {Fa}~\bibnamefont {Wang}}, \bibinfo {author} {\bibfnamefont {Steven~A.}\ \bibnamefont {Kivelson}}, \ and\ \bibinfo {author} {\bibfnamefont {Dung-Hai}\ \bibnamefont {Lee}},\ }\bibfield  {title} {\enquote {\bibinfo {title} {{Nematicity and quantum paramagnetism in FeSe}},}\ }\href {\doibase 10.1038/nphys3456} {\bibfield  {journal} {\bibinfo  {journal} {Nat. Phys.}\ }\textbf {\bibinfo {volume} {11}},\ \bibinfo {pages} {959--963} (\bibinfo {year} {2015})}\BibitemShut {NoStop}%
\bibitem [{\citenamefont {Aroyo}\ \emph {et~al.}(2006{\natexlab{a}})\citenamefont {Aroyo}, \citenamefont {Perez-Mato}, \citenamefont {Capillas}, \citenamefont {Kroumova}, \citenamefont {Ivantchev}, \citenamefont {Madariaga}, \citenamefont {Kirov},\ and\ \citenamefont {Wondratschek}}]{Aroyo:2006}%
  \BibitemOpen
  \bibfield  {author} {\bibinfo {author} {\bibfnamefont {Mois~Ilia}\ \bibnamefont {Aroyo}}, \bibinfo {author} {\bibfnamefont {Juan~Manuel}\ \bibnamefont {Perez-Mato}}, \bibinfo {author} {\bibfnamefont {Cesar}\ \bibnamefont {Capillas}}, \bibinfo {author} {\bibfnamefont {Eli}\ \bibnamefont {Kroumova}}, \bibinfo {author} {\bibfnamefont {Svetoslav}\ \bibnamefont {Ivantchev}}, \bibinfo {author} {\bibfnamefont {Gotzon}\ \bibnamefont {Madariaga}}, \bibinfo {author} {\bibfnamefont {Asen}\ \bibnamefont {Kirov}}, \ and\ \bibinfo {author} {\bibfnamefont {Hans}\ \bibnamefont {Wondratschek}},\ }\bibfield  {title} {\enquote {\bibinfo {title} {Bilbao crystallographic server: I. databases and crystallographic computing programs},}\ }\href {\doibase doi:10.1524/zkri.2006.221.1.15} {\bibfield  {journal} {\bibinfo  {journal} {Zeitschrift für Kristallographie - Crystalline Materials}\ }\textbf {\bibinfo {volume} {221}},\ \bibinfo {pages} {15} (\bibinfo {year} {2006}{\natexlab{a}})}\BibitemShut {NoStop}%
\bibitem [{\citenamefont {Aroyo}\ \emph {et~al.}(2006{\natexlab{b}})\citenamefont {Aroyo}, \citenamefont {Kirov}, \citenamefont {Capillas}, \citenamefont {Perez-Mato},\ and\ \citenamefont {Wondratschek}}]{Aroyo2:2006}%
  \BibitemOpen
  \bibfield  {author} {\bibinfo {author} {\bibfnamefont {Mois~I.}\ \bibnamefont {Aroyo}}, \bibinfo {author} {\bibfnamefont {Asen}\ \bibnamefont {Kirov}}, \bibinfo {author} {\bibfnamefont {Cesar}\ \bibnamefont {Capillas}}, \bibinfo {author} {\bibfnamefont {J.~M.}\ \bibnamefont {Perez-Mato}}, \ and\ \bibinfo {author} {\bibfnamefont {Hans}\ \bibnamefont {Wondratschek}},\ }\bibfield  {title} {\enquote {\bibinfo {title} {{Bilbao Crystallographic Server. II. Representations of crystallographic point groups and space groups}},}\ }\href {\doibase 10.1107/S0108767305040286} {\bibfield  {journal} {\bibinfo  {journal} {Acta Crystallographica Section A}\ }\textbf {\bibinfo {volume} {62}},\ \bibinfo {pages} {115} (\bibinfo {year} {2006}{\natexlab{b}})}\BibitemShut {NoStop}%
\bibitem [{\citenamefont {Cvetkovic}\ and\ \citenamefont {Vafek}(2013)}]{Cvetkovic2013Oct}%
  \BibitemOpen
  \bibfield  {author} {\bibinfo {author} {\bibfnamefont {Vladimir}\ \bibnamefont {Cvetkovic}}\ and\ \bibinfo {author} {\bibfnamefont {Oskar}\ \bibnamefont {Vafek}},\ }\bibfield  {title} {\enquote {\bibinfo {title} {{Space group symmetry, spin-orbit coupling, and the low-energy effective Hamiltonian for iron-based superconductors}},}\ }\href {\doibase 10.1103/PhysRevB.88.134510} {\bibfield  {journal} {\bibinfo  {journal} {Phys. Rev. B}\ }\textbf {\bibinfo {volume} {88}},\ \bibinfo {pages} {134510} (\bibinfo {year} {2013})}\BibitemShut {NoStop}%
\bibitem [{\citenamefont {Suh}\ \emph {et~al.}(2023)\citenamefont {Suh}, \citenamefont {Yu}, \citenamefont {Shishidou}, \citenamefont {Weinert}, \citenamefont {Brydon},\ and\ \citenamefont {Agterberg}}]{Suh2023Sep}%
  \BibitemOpen
  \bibfield  {author} {\bibinfo {author} {\bibfnamefont {Han~Gyeol}\ \bibnamefont {Suh}}, \bibinfo {author} {\bibfnamefont {Yue}\ \bibnamefont {Yu}}, \bibinfo {author} {\bibfnamefont {Tatsuya}\ \bibnamefont {Shishidou}}, \bibinfo {author} {\bibfnamefont {Michael}\ \bibnamefont {Weinert}}, \bibinfo {author} {\bibfnamefont {P.~M.~R.}\ \bibnamefont {Brydon}}, \ and\ \bibinfo {author} {\bibfnamefont {Daniel~F.}\ \bibnamefont {Agterberg}},\ }\bibfield  {title} {\enquote {\bibinfo {title} {{Superconductivity of anomalous pseudospin in nonsymmorphic materials}},}\ }\href {\doibase 10.1103/PhysRevResearch.5.033204} {\bibfield  {journal} {\bibinfo  {journal} {Phys. Rev. Res.}\ }\textbf {\bibinfo {volume} {5}},\ \bibinfo {pages} {033204} (\bibinfo {year} {2023})}\BibitemShut {NoStop}%
\bibitem [{\citenamefont {Abanov}\ \emph {et~al.}(2003)\citenamefont {Abanov}, \citenamefont {Chubukov},\ and\ \citenamefont {Schmalian}}]{Abanov2003Mar}%
  \BibitemOpen
  \bibfield  {author} {\bibinfo {author} {\bibfnamefont {{\relax Ar}.}~\bibnamefont {Abanov}}, \bibinfo {author} {\bibfnamefont {Andrey~V.}\ \bibnamefont {Chubukov}}, \ and\ \bibinfo {author} {\bibfnamefont {J.}~\bibnamefont {Schmalian}},\ }\bibfield  {title} {\enquote {\bibinfo {title} {{Quantum-critical theory of the spin-fermion model and its application to cuprates: Normal state analysis}},}\ }\href {https://www.tandfonline.com/doi/abs/10.1080/0001873021000057123} {\bibfield  {journal} {\bibinfo  {journal} {Adv. Phys.}\ } (\bibinfo {year} {2003})}\BibitemShut {NoStop}%
\bibitem [{\citenamefont {Eschrig}\ and\ \citenamefont {Norman}(2000)}]{Eschrig2000Oct}%
  \BibitemOpen
  \bibfield  {author} {\bibinfo {author} {\bibfnamefont {M.}~\bibnamefont {Eschrig}}\ and\ \bibinfo {author} {\bibfnamefont {M.~R.}\ \bibnamefont {Norman}},\ }\bibfield  {title} {\enquote {\bibinfo {title} {{Neutron Resonance: Modeling Photoemission and Tunneling Data in the Superconducting State of ${{\mathrm{Bi}}_{2}{\mathrm{Sr}}_{2}{\mathrm{CaCu}}_{2}O}_{8+\mathit{\ensuremath{\delta}}}$}},}\ }\href {\doibase 10.1103/PhysRevLett.85.3261} {\bibfield  {journal} {\bibinfo  {journal} {Phys. Rev. Lett.}\ }\textbf {\bibinfo {volume} {85}},\ \bibinfo {pages} {3261} (\bibinfo {year} {2000})}\BibitemShut {NoStop}%
\bibitem [{\citenamefont {Vilk}\ and\ \citenamefont {Tremblay}(1997)}]{Vilk1997Nov}%
  \BibitemOpen
  \bibfield  {author} {\bibinfo {author} {\bibfnamefont {Y.~M.}\ \bibnamefont {Vilk}}\ and\ \bibinfo {author} {\bibfnamefont {A.-M.~S.}\ \bibnamefont {Tremblay}},\ }\bibfield  {title} {\enquote {\bibinfo {title} {{Non-Perturbative Many-Body Approach to the Hubbard Model and Single-Particle Pseudogap}},}\ }\href {\doibase 10.1051/jp1:1997135} {\bibfield  {journal} {\bibinfo  {journal} {J. Phys. I}\ }\textbf {\bibinfo {volume} {7}},\ \bibinfo {pages} {1309} (\bibinfo {year} {1997})}\BibitemShut {NoStop}%
\bibitem [{\citenamefont {Ortenzi}\ \emph {et~al.}(2009)\citenamefont {Ortenzi}, \citenamefont {Cappelluti}, \citenamefont {Benfatto},\ and\ \citenamefont {Pietronero}}]{Ortenzi:2009}%
  \BibitemOpen
  \bibfield  {author} {\bibinfo {author} {\bibfnamefont {L.}~\bibnamefont {Ortenzi}}, \bibinfo {author} {\bibfnamefont {E.}~\bibnamefont {Cappelluti}}, \bibinfo {author} {\bibfnamefont {L.}~\bibnamefont {Benfatto}}, \ and\ \bibinfo {author} {\bibfnamefont {L.}~\bibnamefont {Pietronero}},\ }\bibfield  {title} {\enquote {\bibinfo {title} {Fermi-surface shrinking and interband coupling in iron-based pnictides},}\ }\href {\doibase 10.1103/PhysRevLett.103.046404} {\bibfield  {journal} {\bibinfo  {journal} {Phys. Rev. Lett.}\ }\textbf {\bibinfo {volume} {103}},\ \bibinfo {pages} {046404} (\bibinfo {year} {2009})}\BibitemShut {NoStop}%
\bibitem [{\citenamefont {Bhattacharyya}\ \emph {et~al.}(2020)\citenamefont {Bhattacharyya}, \citenamefont {Bj\"ornson}, \citenamefont {Zantout}, \citenamefont {Steffensen}, \citenamefont {Fanfarillo}, \citenamefont {Kreisel}, \citenamefont {Valent\'{\i}}, \citenamefont {Andersen},\ and\ \citenamefont {Hirschfeld}}]{Bhattacharyya:2020}%
  \BibitemOpen
  \bibfield  {author} {\bibinfo {author} {\bibfnamefont {Shinibali}\ \bibnamefont {Bhattacharyya}}, \bibinfo {author} {\bibfnamefont {Kristofer}\ \bibnamefont {Bj\"ornson}}, \bibinfo {author} {\bibfnamefont {Karim}\ \bibnamefont {Zantout}}, \bibinfo {author} {\bibfnamefont {Daniel}\ \bibnamefont {Steffensen}}, \bibinfo {author} {\bibfnamefont {Laura}\ \bibnamefont {Fanfarillo}}, \bibinfo {author} {\bibfnamefont {Andreas}\ \bibnamefont {Kreisel}}, \bibinfo {author} {\bibfnamefont {Roser}\ \bibnamefont {Valent\'{\i}}}, \bibinfo {author} {\bibfnamefont {Brian~M.}\ \bibnamefont {Andersen}}, \ and\ \bibinfo {author} {\bibfnamefont {P.~J.}\ \bibnamefont {Hirschfeld}},\ }\bibfield  {title} {\enquote {\bibinfo {title} {Nonlocal correlations in iron pnictides and chalcogenides},}\ }\href {\doibase 10.1103/PhysRevB.102.035109} {\bibfield  {journal} {\bibinfo  {journal} {Phys. Rev. B}\ }\textbf {\bibinfo {volume} {102}},\ \bibinfo {pages} {035109} (\bibinfo {year} {2020})}\BibitemShut {NoStop}%
\bibitem [{\citenamefont {Chang}\ \emph {et~al.}(2025)\citenamefont {Chang}, \citenamefont {Backes}, \citenamefont {Lu}, \citenamefont {Gauthier}, \citenamefont {Hashimoto}, \citenamefont {Chen}, \citenamefont {Wen}, \citenamefont {Mo}, \citenamefont {Shen}, \citenamefont {Valent{\ifmmode\acute{\imath}\else\'{\i}\fi}},\ and\ \citenamefont {Pfau}}]{Chang2025Jul}%
  \BibitemOpen
  \bibfield  {author} {\bibinfo {author} {\bibfnamefont {Ming-Hua}\ \bibnamefont {Chang}}, \bibinfo {author} {\bibfnamefont {Steffen}\ \bibnamefont {Backes}}, \bibinfo {author} {\bibfnamefont {Donghui}\ \bibnamefont {Lu}}, \bibinfo {author} {\bibfnamefont {Nicolas}\ \bibnamefont {Gauthier}}, \bibinfo {author} {\bibfnamefont {Makoto}\ \bibnamefont {Hashimoto}}, \bibinfo {author} {\bibfnamefont {Guan-Yu}\ \bibnamefont {Chen}}, \bibinfo {author} {\bibfnamefont {Hai-Hu}\ \bibnamefont {Wen}}, \bibinfo {author} {\bibfnamefont {Sung-Kwan}\ \bibnamefont {Mo}}, \bibinfo {author} {\bibfnamefont {Zhi-Xun}\ \bibnamefont {Shen}}, \bibinfo {author} {\bibfnamefont {Roser}\ \bibnamefont {Valent{\ifmmode\acute{\imath}\else\'{\i}\fi}}}, \ and\ \bibinfo {author} {\bibfnamefont {Heike}\ \bibnamefont {Pfau}},\ }\bibfield  {title} {\enquote {\bibinfo {title} {{Observation of two cascading screening processes in an iron-based superconductor}},}\ }\href {\doibase 10.1038/s43246-025-00881-5} {\bibfield  {journal} {\bibinfo  {journal}
  {Commun. Mater.}\ }\textbf {\bibinfo {volume} {6}},\ \bibinfo {pages} {1--9} (\bibinfo {year} {2025})}\BibitemShut {NoStop}%
\bibitem [{\citenamefont {Zou}\ \emph {et~al.}(2025)\citenamefont {Zou}, \citenamefont {Sihi}, \citenamefont {Oli}, \citenamefont {Roig}, \citenamefont {Agterberg}, \citenamefont {Weinert}, \citenamefont {Li},\ and\ \citenamefont {Mandal}}]{Zou2025Jun}%
  \BibitemOpen
  \bibfield  {author} {\bibinfo {author} {\bibfnamefont {Qiang}\ \bibnamefont {Zou}}, \bibinfo {author} {\bibfnamefont {Antik}\ \bibnamefont {Sihi}}, \bibinfo {author} {\bibfnamefont {Basu~Dev}\ \bibnamefont {Oli}}, \bibinfo {author} {\bibfnamefont {Merc{\ifmmode\grave{e}\else\`{e}\fi}}\ \bibnamefont {Roig}}, \bibinfo {author} {\bibfnamefont {Daniel}\ \bibnamefont {Agterberg}}, \bibinfo {author} {\bibfnamefont {Michael}\ \bibnamefont {Weinert}}, \bibinfo {author} {\bibfnamefont {Lian}\ \bibnamefont {Li}}, \ and\ \bibinfo {author} {\bibfnamefont {Subhasish}\ \bibnamefont {Mandal}},\ }\bibfield  {title} {\enquote {\bibinfo {title} {{Correlation Enhanced Electron-Phonon Coupling in FeSe/SrTiO$_3$ at a Magic Angle}},}\ }\href {\doibase 10.48550/arXiv.2506.22435} {\bibfield  {journal} {\bibinfo  {journal} {arXiv}\ } (\bibinfo {year} {2025}),\ 10.48550/arXiv.2506.22435},\ \Eprint {http://arxiv.org/abs/2506.22435} {2506.22435} \BibitemShut {NoStop}%
\bibitem [{\citenamefont {Ge}\ \emph {et~al.}(2019)\citenamefont {Ge}, \citenamefont {Yan}, \citenamefont {Zhang}, \citenamefont {Agterberg}, \citenamefont {Weinert},\ and\ \citenamefont {Li}}]{Ge2019Apr}%
  \BibitemOpen
  \bibfield  {author} {\bibinfo {author} {\bibfnamefont {Zhuozhi}\ \bibnamefont {Ge}}, \bibinfo {author} {\bibfnamefont {Chenhui}\ \bibnamefont {Yan}}, \bibinfo {author} {\bibfnamefont {Huimin}\ \bibnamefont {Zhang}}, \bibinfo {author} {\bibfnamefont {Daniel}\ \bibnamefont {Agterberg}}, \bibinfo {author} {\bibfnamefont {Michael}\ \bibnamefont {Weinert}}, \ and\ \bibinfo {author} {\bibfnamefont {Lian}\ \bibnamefont {Li}},\ }\bibfield  {title} {\enquote {\bibinfo {title} {{Evidence for d-Wave Superconductivity in Single Layer FeSe/SrTiO3 Probed by Quasiparticle Scattering Off Step Edges}},}\ }\href {\doibase 10.1021/acs.nanolett.9b00135} {\bibfield  {journal} {\bibinfo  {journal} {Nano Lett.}\ }\textbf {\bibinfo {volume} {19}},\ \bibinfo {pages} {2497--2502} (\bibinfo {year} {2019})}\BibitemShut {NoStop}%
\bibitem [{\citenamefont {Yang}\ \emph {et~al.}(2019)\citenamefont {Yang}, \citenamefont {Yan}, \citenamefont {Ma}, \citenamefont {Li},\ and\ \citenamefont {Cen}}]{Yang2019}%
  \BibitemOpen
  \bibfield  {author} {\bibinfo {author} {\bibfnamefont {Ming}\ \bibnamefont {Yang}}, \bibinfo {author} {\bibfnamefont {Chenhui}\ \bibnamefont {Yan}}, \bibinfo {author} {\bibfnamefont {Yanjun}\ \bibnamefont {Ma}}, \bibinfo {author} {\bibfnamefont {Lian}\ \bibnamefont {Li}}, \ and\ \bibinfo {author} {\bibfnamefont {Cheng}\ \bibnamefont {Cen}},\ }\bibfield  {title} {\enquote {\bibinfo {title} {Light induced non-volatile switching of superconductivity in single layer {F}e{S}e on {S}r{T}i{O}$_3$ substrate},}\ }\href {\doibase 10.1038/s41467-018-08024-w} {\bibfield  {journal} {\bibinfo  {journal} {Nature Communications}\ }\textbf {\bibinfo {volume} {10}},\ \bibinfo {pages} {85} (\bibinfo {year} {2019})}\BibitemShut {NoStop}%
\end{thebibliography}%

\newpage

\ifarXiv
    \foreach \x in {1,...,\numbersupplementpages}
    {
        \clearpage
        \includepdf[pages={\x}]{\supplementfilename}
    }
\fi

\end{document}